\input harvmac
\input epsf.tex
\overfullrule=0mm
\newcount\figno
\figno=0
\def\fig#1#2#3{
\par\begingroup\parindent=0pt\leftskip=1cm\rightskip=1cm\parindent=0pt
\baselineskip=11pt
\global\advance\figno by 1
\midinsert
\epsfxsize=#3
\centerline{\epsfbox{#2}}
\vskip 12pt
{\bf Fig.\the\figno:} #1\par
\endinsert\endgroup\par
}
\def\figlabel#1{\xdef#1{\the\figno}}
\def\encadremath#1{\vbox{\hrule\hbox{\vrule\kern8pt\vbox{\kern8pt
\hbox{$\displaystyle #1$}\kern8pt}
\kern8pt\vrule}\hrule}}
\def\appendix#1#2{\global\meqno=1\global\subsecno=0\xdef\secsym{\hbox{#1.}}
\bigbreak\bigskip\noindent{\bf Appendix#1. #2}\message{(#1. #2)}
\writetoca{Appendix {#2}}\par\nobreak\medskip\nobreak}
\def\em{\it}

\def\ZZ{Z\kern -5pt Z}

\Title{T98/078, S98/051}
{{\vbox {
\medskip
\centerline{{\bf Statistical and Dynamical Properties}}
\centerline{{\bf of the Discrete Sinai Model}}
\centerline{{\bf at Finite Times}}
}}}

\bigskip
\centerline{J. Chave\footnote{}{\kern -20pt email: chave@spec.saclay.cea.fr, 
guitter@spht.saclay.cea.fr},}
\centerline{ \it CEA, Service de Physique de l'Etat Condens\'e,}
\centerline{ \it F-91191 Gif sur Yvette Cedex, France}
\medskip
\medskip
\centerline{E. Guitter}
\medskip
\centerline{ \it CEA, Service de Physique Th\'eorique de Saclay,}
\centerline{ \it F-91191 Gif sur Yvette Cedex, France}
\baselineskip=12pt
\vskip .5in
We study the Sinai model for the diffusion of a particle in a one dimensional
quenched random energy landscape. We consider the particular case of discrete
energy landscapes made of random $\pm 1$ jumps on the semi infinite line 
$\ZZ^+$ with a reflecting wall at the origin. We compare the statistical 
distribution of the successive local minima of the energy landscapes, which 
we derive explicitly, with the dynamical distribution of the position of the 
diffusing particle, which we obtain numerically. At high temperature, the 
two distributions match only in the large time asymptotic regime. At low 
temperature however, we find even at finite times a clear correspondence 
between the statistical and dynamical distributions, with additional interesting 
oscillatory behaviors.

\Date{09/98}
\nref\KKS{H. Kesten, M.W. Koslow and F. Spitzer,
{\em Compos. Math.}, {\bf 30}, 145-168, (1975)}
\nref\Sinai{Ya. G. Sinai, {\em Theor. Proba. Appl.},
{\bf 27}, 256-268, (1982)}
\nref\Golun{A.O. Golosov, {\em Soviet Math. Dokl.}, {\bf 28}, 18-22, (1983)}
\nref\Kes{H. Kesten, {\em Physica A}, {\bf 138}, 299-309, (1986)}
\nref\Goltrois{A.O. Golosov, {\em Russ. Math. Surveys}, {\bf 41}, 199-201, 
(1986)}
\nref\Goldeux{A.O. Golosov, {\em Commun. Math. Phys.}, {\bf 92}, 491-506, 
(1984)}
\nref\Bou{J.-P. Bouchaud, A. Comtet, A. Georges and P. Le Doussal,
{\em Ann. Phys.}, {\bf 201}, 285-341, (1990)}
\nref\Bun{A. Bunde, S. Havlin, H.E. Roman, G. Schildt and H.E. Stanley,
{\em J. Stat. Phys.}, {\bf 50}, 1271-1276, (1988)}
\nref\Gras{P. Grassberger and R. Leuverink,
{\em J. Phys. A: Math. Gen.}, {\bf 23}, 773-780, (1990)}
\nref\Lal{L. Laloux and P. Le Doussal, {\em Phys. Rev. E}, {\bf 57}, 6296-6326, 
(1998), (cond-mat/9705249)}
\nref\FLDM{D.S. Fisher, P. Le Doussal and C. Monthus,
{\em Phys. Rev. Lett.}, {\bf 80}, 3539-3542, (1998), (cond-mat/9710270)}
\nref\DGG{P. Di Francesco, O. Golinelli and E. Guitter, {\em Commun. Math. 
Phys.}, {\bf 186}, 1-59, (1997), (cond-mat/9602025)}
\nref\Compte{A. Compte and J.-P. Bouchaud,
{\em J. Phys. A.: Math. Gen.}, {\bf 31}, 6113-6121, (1998), (cond-mat/9801140)}

\newsec{Introduction}

The problem of the diffusion of a particle in a one dimensional quenched 
random energy landscape has been studied for many years [\xref\KKS,\xref\Sinai]. 
If the landscape itself has a random walk statistics, many exact predictions
exist for the {\it long time behavior} of the diffusion process. 
For an unbiased statistics of the potential, i.e in the absence of drift, 
the average distance $\langle x\rangle$ traveled by the particle grows 
very slowly at large times, with the Sinai scaling behavior 
$\langle x\rangle \sim \left(T{\ln}(t)\right)^2$, where $T$ is the 
temperature \Sinai . 
The asymptotic probability distribution is universal in the scaling variable 
$x/\left(T{\ln}(t)\right)^2$ and its precise form is known exactly 
for a single diffusing particle [\xref\Golun-\xref\Goltrois]. Moreover, 
the thermal 
dispersion (or the distance between two diffusing particles in the same 
random potential) remains finite at large times \Goldeux. The physical picture
underlying these results is simply that of a {\it localization} of the 
diffusing particle in the deepest energy minimum which it can reach at time $t$ 
by passing larger and larger energy barriers [\xref\Goldeux-\xref\Bou].

These large time asymptotic predictions were tested numerically by computing 
the {\it exact} probability distribution for the position $x$ of the particle
at finite but large enough times $t$ for energy landscapes drawn at random, 
and then averaging over a large enough sample of such landscapes 
[\xref\Bun-\xref\Lal]. One should notice that the Sinai scaling 
$\langle x \rangle \sim (T \ln(t))^2$ defines a very slow diffusive process. 
For such a process, the long time behavior is expected to
be reached in practice after extremely long transients, especially at 
low temperature. Such times can be out of reach in practical situations or
in finite time numerical simulations. This probably explains why the asymptotic 
predictions are not fully recovered in \Bun .
In any case, very little is known about the finite time behavior of the model
or about the approach to the asymptotic regime.

In practice, the localization of the particle in the deepest minimum has
the nice consequence that some of the {\it dynamical} properties of the 
diffusion process can be deduced directly from the corresponding 
{\it statistical}
properties of the minima of the random potential.
More precisely, the underlying idea is to assume 
that, at times of order $t$, the particle is localized in the deepest 
minimum of the energy landscape which it could reach by passing all 
the energy barriers $\Delta E$ of height less or equal to 
$\Gamma\sim T{\ln}(t)$. The time dependence of various quantities in 
the dynamical process can then be obtained from the equilibrium {\it statistics} 
of the energy minima and its dependence in the highest passable energy 
barrier $\Gamma$. 
This idea has been implemented recently to develop a new Real Space 
Renormalization 
Group (RSRG) approach for the Sinai model \FLDM . The RSRG formalism consists in
performing a suitable decimation of the barriers with height less than $\Gamma$ 
to obtain universal Renormalization Group equations for the variation with 
increasing 
$\Gamma$ of the effective energy landscape distribution seen by the particle 
at the time scale $t$. Long time properties are obtained from the 
$\Gamma \to \infty$ fixed point of these RSRG equations. This technique 
allows to recover the universal asymptotic probability distribution of a
single particle for various boundary conditions. It also allows for the 
prediction of asymptotic two-time or two-particle correlations.

Following the above ``statistical picture'' leading to the exact asymptotic 
predictions, one may wonder whether this picture could also help to investigate 
shorter times or very low temperatures. 
In this paper, we test the connection between the dynamical properties of 
the diffusion process and the statistics of the minima of the energy 
landscapes for finite times. We use a particular distribution of the disorder 
where the energy landscapes are made of successive random $\pm 1$ increments, 
i.e. have the 
statistics of a {\it discrete} random walk. The reason for this choice is 
twofold. On the one hand, we can easily derive {\it explicit} laws for the 
statistics of minima at arbitrary (even small) $\Gamma$. On the other hand, 
this is particularly adapted to numerical simulations of the dynamical 
process. As we shall see, our results corroborate the statistical 
picture, not only at large $\Gamma$ where we recover the universal 
distributions expected for general random energy landscapes with a random
walk statistics, but also at small $\Gamma$ {\it and} low temperature, 
where the discrete nature of the energy landscapes is sensible and yields 
interesting behaviors. 

The paper is organized as follows. In Section 2, we briefly describe the 
particle diffusion process and the distribution for the random energy 
landscapes. In Section 3, we derive explicit formulas for the statistics 
of energy minima in the presence of energy barriers of arbitrary scale 
$\Gamma$. Section 4 presents numerical results for the dynamical diffusion 
process. These results are compared with the predictions of the statistical 
approach of Section 3.  We gather our conclusions in Section 5.
 
\newsec{Diffusion of a particle in a discrete random energy landscape}
\fig{A discrete random energy landscape $E(x)$. A reflecting wall prevents
the particle from exploring the negative $x$ region. The particle jumps
to the left or to the right with the probabilities given by equation (2.2).}
{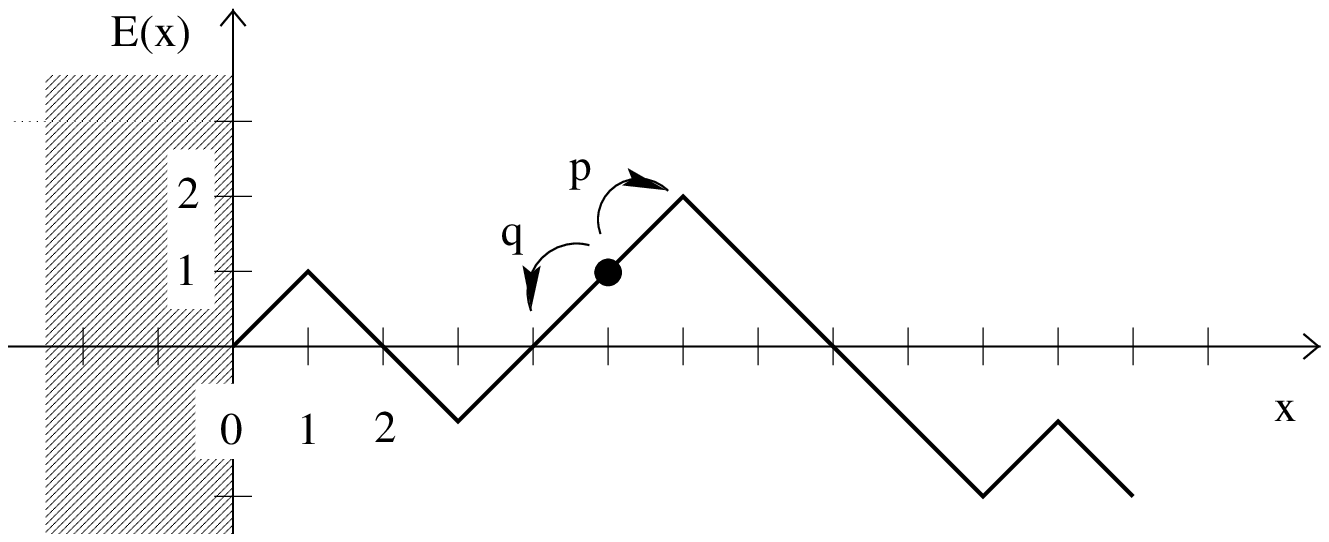}{7.cm}
\figlabel\model
We consider the following discrete distribution for the quenched energy 
landscapes. On each site $x=0,1,2,...$ 
labeled by a non negative integer, we associate a random energy variable 
$E(x)$ such that (see Figure \model )
\eqn\height{\eqalign{&E(0)=0 \cr & \Delta E(x)=E(x+1)-E(x)=\pm 1 \cr}}
The sign $\pm 1$ of the increment is drawn at random between successive sites 
with even probability $1/2$, i.e. we introduce no drift in the problem. 
With the choice \height , the values of
the energy are limited to integers, meaning that we have
implicitly fixed an underlying energy scale. 
We can view $E(x)$ as the ``height" at site $x$ of a discrete random walk 
describing our energy landscape.  We finally set $E(-1)=+\infty$, i.e. 
put an infinite reflecting wall on the negative side of the origin, while
we keep a free boundary condition at $x=+\infty$.
This choice of boundary condition is clearly not crucial but makes our 
analysis simpler since in this case, deeper minima always appear at
increasing values of $x$. 

For a given realization of the quenched random energy landscape,
we consider the following discrete time dynamical process:
\item{-} We start at time $t=0$ at position $x=0$ with energy $E(0)=0$.
\item{-} Given the position $x$ with energy $E(x)$ at time $t$, the
position at time $t+1$ is 
\eqn\jum{\left\{\eqalign{x+1 &\hbox{ with probability }
p_x\equiv {e^{-\beta\Delta E(x)} \over e^{-\beta\Delta E(x)}
+e^{\beta\Delta E(x-1)}} \cr 
x-1 &\hbox{ with probability } 
q_x\equiv {e^{\beta\Delta E(x-1)} \over e^{-\beta\Delta E(x)}
+e^{\beta\Delta E(x-1)}} \cr 
}\right.} 
with $\Delta E(x)$ as in \height\ and $\beta=1/T$ the inverse 
temperature. 
\par
With the choice \height\ for the variation of the energy
on neighboring sites, the ratio $p_x/q_x$ is equal to $1$ if the two sites
$x+1$ and $x-1$ have the same energy, or to $\exp(\pm 2/T)$ if they have
different energies. Note that the case $x=0$ is special with
$p_0=1-q_0=1$. Note also that the particle has to move to
one of its neighboring sites at each time step, i.e. we do not allow 
the particle to remain at the same site. As a consequence, the particle 
occupies sites with even position at even times and sites with odd position
at odd times. This effect results in ``residual fluctuations"
even at $T=0$. The above model is very simple since it depends on only 
one parameter, the temperature $T$. We study its behavior numerically 
in Section 4 but let us first derive explicit laws for the minima of energy
landscapes with the discrete random walk statistics \height .

\newsec{Statistics of energy minima}

In this section, we forget for a while the dynamical process and focus
on the statistics of the successive minima of the random energy landscapes 
which can be reached in the presence of increasingly passable energy barriers. 
Following \FLDM , we consider for any given realization of the energy 
landscape the position $x=x_{\rm min}(\Gamma)$ of the {\it deepest} minimum 
of this landscape which can be reached, starting from the wall at the 
origin $x=0$, by passing through energy barriers of height {\it less or 
equal} to a fixed scale $\Gamma$. Here $x_{\rm min}$ and $\Gamma$ are 
non negative integers. For each $\Gamma$, we compute the 
probability $p_\Gamma(x)$ that $x_{\rm min}(\Gamma)$ equals $x$, 
obtained by averaging the quantity $\delta_{x,x_{\rm min}(\Gamma)}$ 
over all realizations of the energy landscape. Note that, for a given 
realization, the deepest reachable minimum is not necessarily unique 
but can be degenerate. 
We will thus consider three different probabilities $p^{(1)}_\Gamma(x)$, 
$p^{(2)}_\Gamma(x)$ and $p^{(3)}_\Gamma(x)$, corresponding respectively 
to the following prescriptions in case of degeneracy: 
\item{(1)} we keep only the closest minimum to the origin, 
\item{(2)} we keep only the furthest minimum from the origin, 
\item{(3)} we take the average over all the degenerate minima, 
i.e. consider the quantity $(1/k)\sum_{i=1}^k \delta_{x,x_{\rm min}^{(k)}}$ 
for $k$ degenerate minima at position $x_{\rm min}^{(k)}$.\par
\noindent  We shall give below explicit formulas for the generating functions
\eqn\gener{{\cal P}^{(i)}_\Gamma(z)=\sum_{x=0}^{\infty}p^{(i)}_\Gamma(x) z^x
\ \ i=1,2,3}

\subsec{Random walk in a strip of height $\Gamma$}

Before we proceed, we need to evaluate the probability 
$d_\Gamma(x)$ for a random walk starting at $E(0)=0$ to have its $x$ first 
steps inside the strip $0\le E \le \Gamma$ and its $(x+1)$th step at
$E(x+1)=-1$, i.e. the probability for the walk to leave the strip for
the first time just after position $x$ and downwards. We recall 
here the explicit form of the generating function $D_\Gamma(z) =\sum_x 
d_\Gamma(x) z^x$. In the following, we will always assume that the $(x+1)$th 
step points downwards, which in practice occurs with probability $1/2$, 
i.e. we will calculate $2D_\Gamma(z)$ instead of $D_\Gamma(z)$. 
\fig{For a strip of height $\Gamma=1$, the random walk is a series of 
$E=0 \to 1\to 0$ two-step-sequences. To obtain $2D_\Gamma(z)$, 
each two-step-sequence must be dressed with a factor $2D_{\Gamma-1}(z)$
in the middle.}{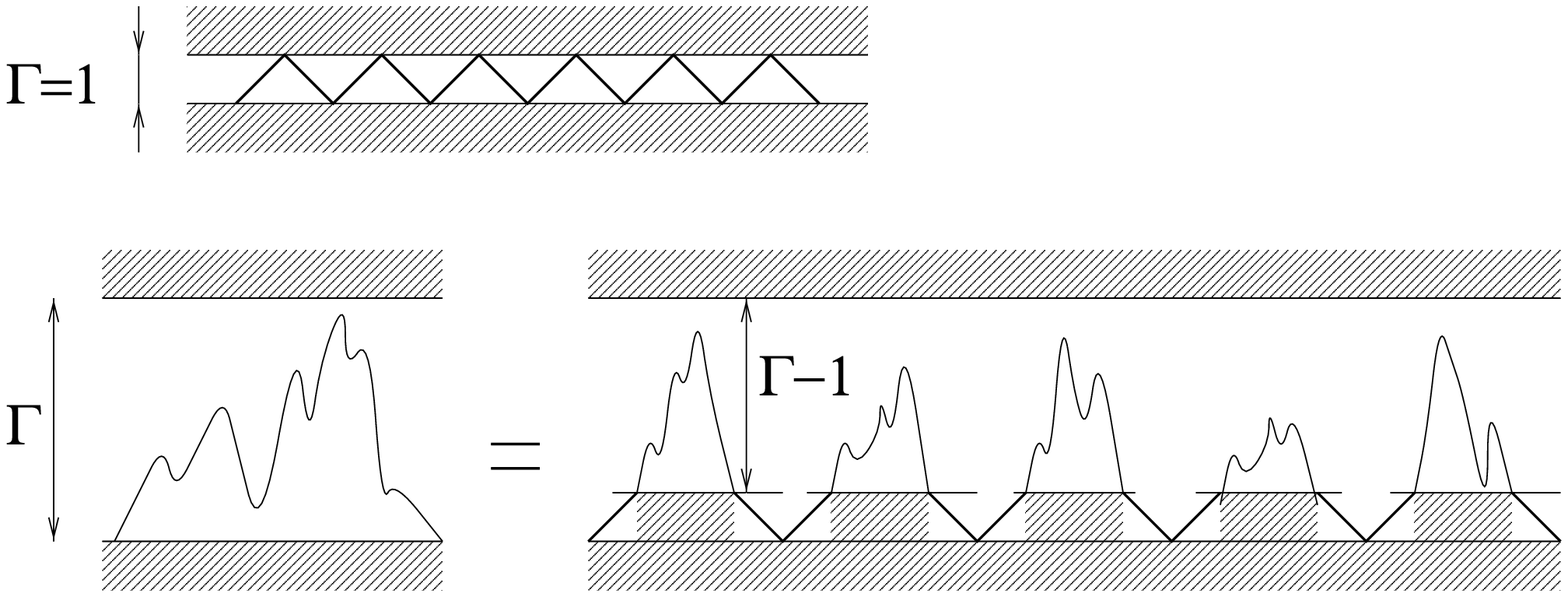}{9.cm}
\figlabel\recurrence
The function $D_\Gamma(z)$ satisfies the recursion relation:
\eqn\recur{D_\Gamma(z)={1\over 2-z^2 D_{\Gamma-1}(z)}}
with the initial condition $2D_0(z)=1$ (for $\Gamma=0$, the first exit 
from the zero-width strip occurs necessarily just after $x=0$).
This relation can be understood as follows (see Figure \recurrence):
consider the case $\Gamma=1$, for which the random walk is necessarily
of even length $x$ and made of the repetition of $x/2$ elementary 
two-step-sequences of the type ($E=0\to 1\to 0$). 
Each two-step-sequence comes 
with a factor $z^2$ in the generating function and occurs with a probability 
$(1/2)^2$, leading to,
\eqn\Done{2D_1(z)=\sum_{x=0\atop x\rm even}^\infty \left({z^2\over 2^2}
\right)^{x\over 2}= {1\over 1-\left({z\over 2}\right)^2}}
in agreement with \recur\ and $D_0(z)=1/2$.
The generating function $2D_\Gamma(z)$ can be obtained in the same
way by replacing the weight $(z/2)^2$ for the elementary two-step-sequence
by a weight $(z/2)^2\times 2D_{\Gamma-1}(z)$, corresponding to inserting
between the two steps $E=0\to 1$ and $E=1\to 0$ an arbitrary sequence 
of steps limited to the strip of width $\Gamma-1$, $1\le E\le \Gamma$
(see Figure \recurrence).
Making this replacement in \Done\ directly leads to \recur .
 
{}From \recur, we can write the generating function $D_\Gamma(z)$ as
the ratio:
\eqn\DPP{D_\Gamma(z)={P_\Gamma(z)\over P_{\Gamma+1}(z)}}
where $P_\Gamma(z)$ is an even polynomial of degree $2[\Gamma/2]$
in the variable $z$. Indeed, writing 
$D_\Gamma(z)=P_\Gamma(z)/ Q_{\Gamma}(z)$, \recur\ gives
\eqn\ratio{{P_\Gamma(z)\over Q_{\Gamma}(z)}={1\over 2- z^2
{P_{\Gamma-1}(z)\over Q_{\Gamma-1}(z)}}={Q_{\Gamma-1}(z)
\over 2Q_{\Gamma-1}(z)-z^2P_{\Gamma-1}(z)}} 
leading to recursion relations 
\eqn\recurPQ{\eqalign{P_\Gamma(z)&=Q_{\Gamma-1}(z)\cr Q_\Gamma(z)
&= 2Q_{\Gamma-1}(z)-z^2P_{\Gamma-1}(z)\cr}}
Eliminating $Q_\Gamma$, we get the announced result \DPP , 
with, moreover, the recursion relation
\eqn\recurP{P_{\Gamma+1}(z)-2P_{\Gamma}(z)+z^2P_{\Gamma-1}(z)=0}
which determines $P_\Gamma(z)$ from the initial values $P_0(z)=1$ 
and $P_1(z)=2$.
Setting 
\eqn\PtoU{P_\Gamma(z)=z^\Gamma U_\Gamma\left({2\over z}\right) 
\leftrightarrow U_\Gamma(y)=\left({y\over 2}\right)^\Gamma P_\Gamma\left(
{2\over y}\right)}
the above recursion relation transforms into
\eqn\recurP{U_{\Gamma+1}(y)-yU_{\Gamma}(y)+U_{\Gamma-1}(y)=0}
with $U_0(y)=1$ and $U_1(y)=y$. The solution of this recursion is given by
the well known Chebishev polynomials $U_\Gamma(z)$, characterized by
\eqn\Chebi{U_\Gamma(2{\rm cosh}(t))={{\rm  sinh}\left((\Gamma+1)t\right)
\over {\rm sinh}(t)}}
We thus get the explicit form:
\eqn\Pform{P_\Gamma\left({1\over {\rm cosh}(t)}\right)=
{1\over ({\rm cosh}(t))^\Gamma}{{\rm  sinh}\left((\Gamma+1)t\right)
\over {\rm sinh}(t)}}
and the explicit formula:
\eqn\Dform{D_\Gamma\left(z={1\over {\rm cosh}(t)}\right)={\rm cosh}(t)
{{\rm  sinh}\left((\Gamma+1)t\right) \over {\rm sinh}\left((\Gamma+2)
t\right)}}
The relation between $D_\Gamma(z)$ and the Chebishev polynomials is also
derived in \DGG\ in the context of the Temperley-Lieb algebra.
In the limit $t\to 0$, we get the particular result
\eqn\Dzone{D_\Gamma(1)={\Gamma+1\over \Gamma +2}}
which is the probability for the walk to leave the strip $0\le E\le \Gamma$
downwards at $E=0$ rather that upwards at $E=\Gamma$.
In the limit $\Gamma\to \infty$, we have
\eqn\Dgammainf{D_\Gamma(z)\buildrel {\Gamma\to \infty}\over
\rightarrow {1-\sqrt{1-z^2}\over z^2}}
which, up to normalizations, is the generating function of Catalan numbers,
counting closed walks on the semi-infinite line $E\ge 0$, as it should.

\subsec {Explicit forms of ${\cal P}^{(i)}_\Gamma(z)$}

We now come to the evaluation of the generating functions
${\cal P}^{(i)}_\Gamma(z)$ defined by \gener . Given an
arbitrary landscape of infinite length and for a given $\Gamma$, 
we have to look for the deepest minimum which can be reached by passing 
barriers of height less or equal to $\Gamma$. To do so, 
we decompose the random walk defining the landscape according
to the following algorithm:
\item{1/} Starting at $x_0=0$ and $E(x_0)=0$, the walk has a first part 
entirely included in the strip of width $\Gamma$, $0\le E \le \Gamma$.
In this part, the last return at $E=0$ occurs at
some position $x_1$ (possibly $0$). This defines a first building block
of size $x_1$.
\item{2/} Since $x_1$ is the last return at $E=0$ within the strip of size
$\Gamma$, two situations may occur for the remaining part of the walk: 
\itemitem{2-1/} Either the walk reaches $E=\Gamma+1$ without returning to $E=0$,
i.e. the walk leaves the strip upwards. 
In this case, we have reached a barrier too high to be passed and we stop 
the process.
\itemitem{2-2/} Or $E(x_1+1)=-1$, i.e. the walk leaves the strip downwards. 
In this case we repeat the process 1, i.e 
starting at position $x_1+1$ with $E(x_1+1)=-1$, we consider the following 
part of the walk entirely in the strip of height $\Gamma$,
$-1\le E\le \Gamma-1$ and look for the position $x_2$ of the
last return to $E=-1$ within this strip, defining a second block of 
size $x_2-x_1-1$. We repeat the process until it is stopped by 
a too high barrier. The number ${\cal E}$ of building blocks is simply related 
to the height $1-{\cal E}$ of the deepest minimum which could be reached. 
The position of this minimum can be obtained by properly summing the sizes 
of the blocks.

\fig{A schematic picture of the decomposition of random walks into building
blocks. The number ${\cal E}$ of building blocks is related to the
height $1-{\cal E}$ of the deepest minimum reached without passing a barrier of
size greater than $\Gamma$. For the different building blocks, we have also 
indicated the weights obtained by the average over the random walks
to reconstruct the appropriate generating function ${\cal P}_\Gamma^{(i)}$. 
These weights depend on the precise prescription for degenerate minima: 
(1) keep the closest minimum to the origin; (2) keep the furthest minimum 
from the origin; (3) average over the minima.}{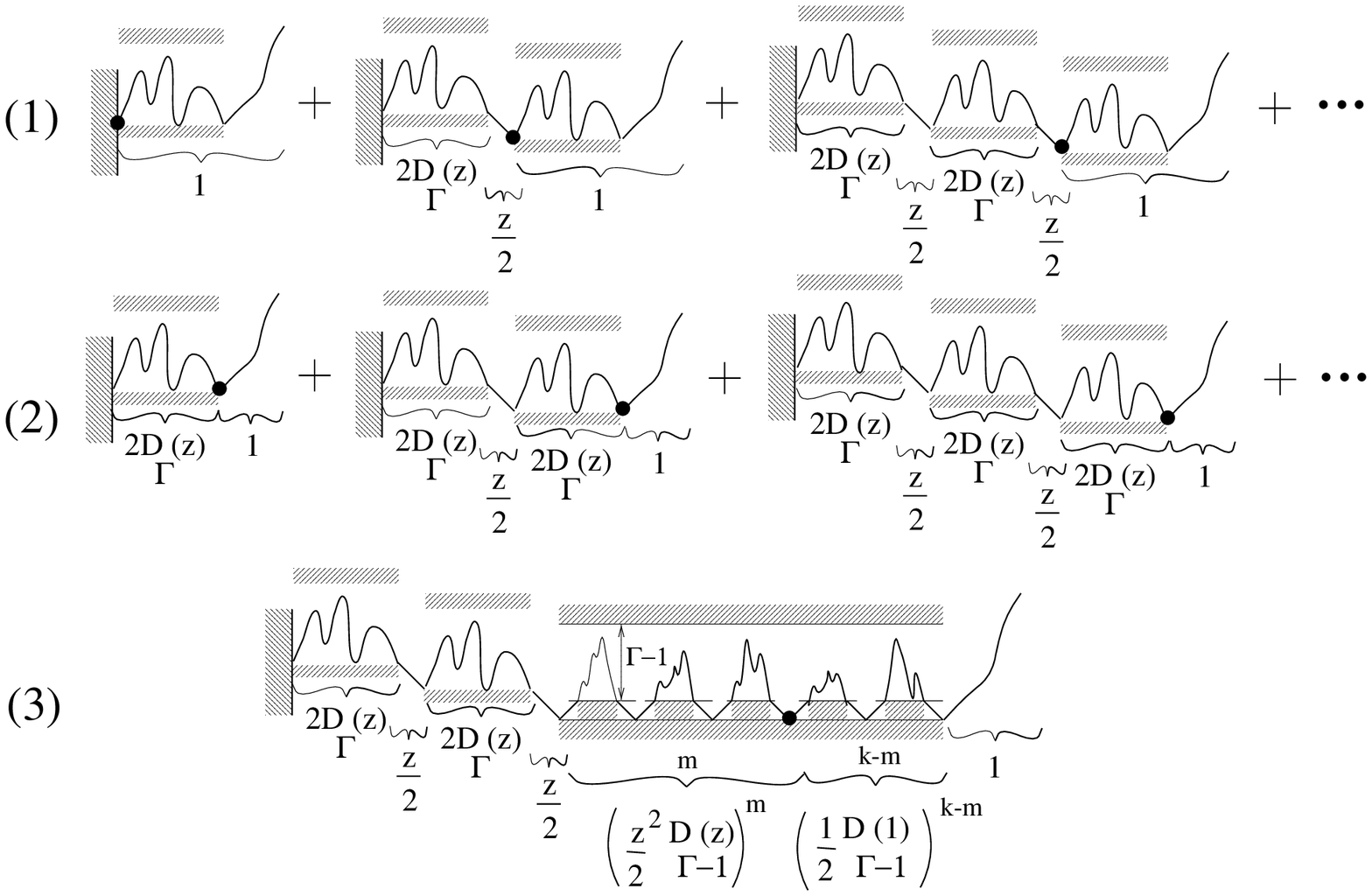}{13.cm}
\figlabel\schema

\noindent  This algorithm is illustrated in Figure \schema. To compute
${\cal P}^{(i)}_\Gamma(z)$, we simply have to assign a weight $2D_\Gamma(z)$
per building block and a factor $(z/2)$ for each descending bond between 
building blocks.
The different prescriptions in case of  degenerate minima concern only the last
building block. In the case $(1)$ of the closest minimum to the origin, the size
of this block does not contribute to the position $x_{\rm min}$ of the minimum.
This block thus comes with a factor $1$. Summing over the number
${\cal E}$ of blocks, we deduce that ${\cal P}^{(1)}_\Gamma(z)$ is proportional
to (see Figure \schema-1):
\eqn\propone{{\cal P}^{(1)}_\Gamma(z)\propto 1+2D_\Gamma(z){z\over 2}+
\left(2D_\Gamma(z){z\over 2}\right)^2+\ldots={1\over 1-zD_\Gamma(z)}}
The normalization factor is fixed by demanding that ${\cal P}^{(1)}_\Gamma(1)=1$
since, for $z=1$, the generating function gives the probability to find
the minimum at some arbitrary position $x$.
We thus get:
\eqn\resone{{\cal P}^{(1)}_\Gamma(z)={1-D_\Gamma(1)\over 1-zD_\Gamma(z)}}
Note that from the previous subsection, we can also understand the prefactor 
$1-D_\Gamma(1)$ as the probability for the random walk to eventually 
leave the strip of size $\Gamma$ upwards rather than going downwards.
For the prescription (2), where we choose the furthest minimum from the origin,
we have an extra factor $2D_\Gamma(z)$ for the last building block.
We thus get in this case (see Figure \schema-2):
\eqn\proptwo{{\cal P}^{(2)}_\Gamma(z)\propto 2D_\Gamma(z)+2D_\Gamma(z){z\over 2}
2D_\Gamma(z)+ \left(2D_\Gamma(z){z\over 2}\right)^22D_\Gamma(z)+\ldots
={2D_\Gamma(z)\over 1-zD_\Gamma(z)}}
and:
\eqn\restwo{{\cal P}^{(2)}_\Gamma(z)={D_\Gamma(z)\over D_\Gamma(1)}\times
{1-D_\Gamma(1)\over 1-zD_\Gamma(z)}}
The difference between the two cases (1) and (2) is thus simply a prefactor
$D_\Gamma(z)/D_\Gamma(1)$ coming from the last building block.  
This factor is nothing but the generating function $\sum_x e_\Gamma(x)z^x$ 
associated with the probability $e_\Gamma(x)$ to find the furthest minimum 
at a distance $x$ from the closest minimum. 

The prescription (3), where we average over all minima in the case of degeneracy 
is more subtle. Here again, it concerns only the weight of last building
block and leads to a different prefactor. Suppose we have $k+1$ 
degenerate minima, each minimum
will receive a weight $1/(k+1)$. For the $(m+1)$'th minimum among the 
$k+1$, we have to assign a factor $z$ only to the elementary steps of the walk
which are to the left of the minimum while the steps which are to the right 
do not contribute to $x_{\rm min}^{(m+1)}$ and receive a factor $1$ instead. 
Still the right part of the block is important to guaranty that the total 
number of minima is precisely $k+1$.
The generating function between successive minima is thus simply 
$2D_{\Gamma-1}(z)(z/2)^2$ on the left side of the chosen minimum and
$2D_{\Gamma-1}(1)(1/2)^2$ on its right side (see Figure \schema-3). The
prefactor associated with the last block is now proportional to:
\eqn\prefave{\eqalign{\sum_{k=0}^{\infty}{1\over k+1}&
\left\{\sum_{m=0}^{k}\left({z^2D_{\Gamma-1}(z)\over 2}\right)^m
\left({D_{\Gamma-1}(1)\over 2}\right)^{k-m}\right\}\cr =&
{1\over {z^2D_{\Gamma-1}(z)\over 2}-{D_{\Gamma-1}(1)\over 2}}
{\ln}\left({1-{z^2D_{\Gamma-1}(z)\over 2} \over 1-
{D_{\Gamma-1}(1)\over 2}}\right)\cr
=&{2D_\Gamma(z)D_\Gamma(1)\over D_\Gamma(z)-D_\Gamma(1)}
{\ln}\left({D_\Gamma(z)\over D_\Gamma(1)}\right)\cr}}
where we have used \recur . We finally get
\eqn\resthree{{\cal P}^{(3)}_\Gamma(z)={D_\Gamma(z)\over D_\Gamma(z)
-D_\Gamma(1)}{\ln}\left({D_\Gamma(z)\over D_\Gamma(1)}\right)\times
{1-D_\Gamma(1)\over 1-zD_\Gamma(z)}}
by adjusting the normalization to guaranty that ${\cal P}^{(3)}_\Gamma(1)=1$.
Note that, as for \restwo , the prefactor involves only a function of
$D_\Gamma(z)/D_\Gamma(1)$.

The formulas \resone-\resthree\ can be made more explicit by use of
\Dform , leading to
\eqn\Ptwoform{\eqalign{{\cal P}^{(1)}_\Gamma\left(z={1\over {\rm cosh}(t)}
\right) & ={1\over \Gamma+2}\times {{\rm  sinh}\left((\Gamma+2)t\right) 
\over {\rm sinh}\left((\Gamma+2) t\right)- {\rm  sinh}\left((\Gamma+1)t\right)
}\cr 
{\cal P}^{(2)}_\Gamma\left(z={1\over {\rm cosh}(t)}\right)
& ={1\over \Gamma+1}\times {{\rm cosh}(t) \ {\rm  sinh}\left((\Gamma+1)t\right) 
\over {\rm sinh}\left((\Gamma+2) t\right)- {\rm  sinh}\left((\Gamma+1)t\right)
}\cr 
{\cal P}^{(3)}_\Gamma\left(z={1\over {\rm cosh}(t)}\right)
& ={{\rm cosh}(t) \ {\rm  sinh}\left((\Gamma+1)t\right) 
\over {\rm sinh}\left((\Gamma+2) t\right)- {\rm  sinh}\left((\Gamma+1)t\right)}
\ln\left[ {\Gamma+2\over \Gamma+1}{{\rm cosh}(t) \ {\rm sinh}\left((\Gamma+1)t
\right) \over {\rm sinh}\left((\Gamma+2)t\right)}\right]
\cr & \ \ \times
{{\rm sinh}\left((\Gamma+2)t\right)\over
(\Gamma+2){\rm cosh}(t)\ {\rm sinh}\left((\Gamma+1)t\right)
-(\Gamma+1){\rm sinh}\left((\Gamma+2)t\right)}\cr
}}
These expressions constitute our central result describing the statistics
of minima at finite $\Gamma$ for the particular distribution of landscapes
that we study.

At large $\Gamma$, an appropriate scaling limit can be obtained by 
a suitable scaling of $z$:
\eqn\scallimit{z=1-{s\over \Gamma^2} \ \ \ i.e. \ \ \ t={\sqrt{2s}\over \Gamma}
+{\cal O}\left({1\over \Gamma^3}\right)}
This corresponds to setting 
\eqn\scalorx{x=\Gamma^2\, l}
in the original probabilities $p_\Gamma^{(i)}(x)$. 
In this limit, the three generating functions ${\cal P}_\Gamma^{(i)}$ 
tend to the same limiting distribution ${\cal P}_\infty(s)$, equal to:
\eqn\limdist{{\cal P}_\infty(s) ={{\rm tanh}(\sqrt{2s})\over \sqrt{2s}}}
This distribution is now the Laplace transform 
\eqn\laptr{{\cal P}_\infty(s)=\int_0^\infty dl\, {p}_\infty(l)e^{-sl}} of
the limiting probability density
\eqn\probdens{{p}_\infty(l)=\lim_{\Gamma\to \infty}\Gamma^2
\,p^{(i)}(x=\Gamma^2l) \quad i=1,2,3}
{}From \limdist , we get
\eqn\limdens{{p}_\infty(l)=\sum_{k=0}^\infty e^{-{(2k+1)^2\over 8}\pi^2l}}
The formulas \limdist\ and \limdens\ are actually more general and valid 
for all the distributions of the energy landscape which have a random
walk statistics, i.e. tend to a Brownian motion distribution in the continuum. 
These laws can be derived directly by use of the RSRG formalism 
developed in \FLDM . They appear there as the large $\Gamma$ fixed point of 
RSRG equations describing the flow with $\Gamma$ of similar renormalized 
probability distributions. Here, we have made explicit the whole $\Gamma$ 
dependence of these probabilities in the special case of an energy 
landscape made of discrete $\pm 1$ increments.

The nice feature of the formulas \limdist\ and \limdens\ 
is that they are precisely those derived in \Golun\ for the limiting
probability distribution of the dynamical process itself in its large time 
scaling regime. In this case, the natural scaling variable is 
$l\equiv \sigma^2x/\ln^2(t)$, where $\sigma=2/T$ is directly related to the
expectation value of $\ln^2(p_x/1-p_x)$. We thus recover here the 
asymptotic equivalence between the statistics of minima and the
dynamical process, with the precise correspondence $\Gamma=\ln(t)/\sigma$.
Such a relation, valid in principle only for large times, will be tested 
in Section 4 at shorter times.  

Finally, let us mention that a similar computation of the generating
function at finite $\Gamma$ can be performed for landscapes without 
wall at the origin. The treatment of the degeneracies is however 
more involved in this case with even more different possible prescriptions. 
Still, the limiting distribution at large $\Gamma$ is independent of 
the chosen prescription.
One can also introduce a drift in the problem when drawing the random 
landscapes. For instance, choosing in \height\ $\Delta E(x)=1$ with 
probability $p$ and $\Delta E(x)=-1$ with probability $q=1-p$, we get
for, say, ${\cal P}^{(1)}_\Gamma$ the expression
\eqn\bias{{\cal P}_\Gamma^{(1)}(z)={1-2\,q\, D_\Gamma\left(\sqrt{4pq}\right)
\over 1-2\, q\, z\, D_\Gamma\left(\sqrt{4pq}z\right)}} 
Different regimes are obtained according to sign of $p-q$, and to whether 
$p-q$ is of order one (strong bias) or of order $1/\Gamma$ (weak bias).
\medskip
To end this section, let us evaluate the first correction 
to ${\cal P}_\infty(s)$ by further expanding the formulas for
${\cal P}_\Gamma^{(i)}$ in powers of $1/\Gamma$. We get
\eqn\corrscal{{\cal P}_\Gamma^{(i)}\left(z=1-{s\over \Gamma^2}\right)=
\left(1-{a^{(i)}\over \Gamma}\right){{\rm tanh}(\sqrt{2s})\over \sqrt{2s}}
+{a^{(i)}\over \Gamma}\left( 1-{3\over 2a^{(i)}}{\rm tanh}^2(\sqrt{2s})\right)
+{\cal O}\left({1\over \Gamma^2}\right)}
with $a^{(1)}=2$, $a^{(2)}=1$ and $a^{(3)}=3/2$.
The above $1/\Gamma$ corrections are probably not universal, i.e. they 
depend on our particular choice for the statistics of landscapes.
Still, the $1/\Gamma$ correction to the {\it relative} differences 
$({\cal P}_\Gamma^{(2)}-{\cal P}_\Gamma^{(1)})/{\cal P}_\Gamma^{(1)}$
and $({\cal P}_\Gamma^{(3)}-{\cal P}_\Gamma^{(1)})/{\cal P}_\Gamma^{(1)}$ 
should be universal.  Indeed, they involve the proportionality factors between 
the different ${\cal P}_\Gamma^{(i)}$'s, i.e. the factor $D_\Gamma(z)/D_\Gamma(1)$ 
in \restwo , or the factor $D_\Gamma(z)/(D_\Gamma(z)
-D_\Gamma(1))\ln(D_\Gamma(z)/D_\Gamma(1))$ in \resthree . 
As we already mentioned, $D_\Gamma(z)/D_\Gamma(1)$ is the generating function for 
the probability $e_\Gamma(x)$ to have the furthest minimum at a distance $x$ from 
the closest. The proportionality factors above thus concern 
the {\it relative} distance between the different degenerate minima,
and ignore their absolute position. They are the important statistical
quantities to be used when one is interested in the localization property
of the dynamical process. If we insist to impose the scaling \scallimit\ 
appropriate to absolute positions, we get, for $z=1-s/\Gamma^2$
\eqn\mauvaisscal{\eqalign{
{D_\Gamma(z)\over D_\Gamma(1)}&=1+{1\over \Gamma}\left(1-\sqrt{2s}\,{\rm coth}
(\sqrt{2s})\right) +{\cal O}\left({1\over \Gamma^2}\right)\cr
{D_\Gamma(z)\over D_\Gamma(z)-D_\Gamma(1)}\ln\left({D_\Gamma(z)\over
D_\Gamma(1)}\right)&=1+{1\over \Gamma}\left({\displaystyle{1-\sqrt{2s}\,{\rm coth}
(\sqrt{2s})}\over 2}\right) +{\cal O}\left({1\over \Gamma^2}\right)\cr
}}
At large $\Gamma$, the factors \mauvaisscal\ above tend to one. This does
not mean that degenerate minima disappear in this limit. Indeed,
one can easily compute the probability for having exactly 
$k+1$ degenerate minima, equal to: 
\eqn\probdeg{\left(1-{D_{\Gamma-1}(1)\over 2}\right)\left({D_{\Gamma-1}(1)
\over 2}\right)^k={\Gamma+2\over 2(\Gamma+1)}\left({\Gamma\over 2(\Gamma+1)}
\right)^k}
Degeneracies do thus exist even at large $\Gamma$ where the
above expression tends to $(1/2)^{k+1}$. 
However, most of these degeneracies occur at short distances, and not
at distances of order $\Gamma^2$. Relative distances of order $\Gamma^2$ 
are found only with a probability of order $1/\Gamma$. The $1/\Gamma$
corrections above correspond to situations with exactly {\it two} 
degenerate minima at a distance of order $\Gamma^2$, as corroborated by 
the fact that averaging over minima gives half the value obtained by 
keeping the furthest minimum. Situations with three or more degenerate 
minima would contribute to higher orders in $1/\Gamma$. 
At large $\Gamma$, the scaling \scallimit\ is therefore not appropriate 
to deal with the relative position of degenerate minima. As noticed in 
\Dgammainf , a non trivial large $\Gamma$ exists for 
$D_\Gamma(z)/D_\Gamma(1)$ {\it without} rescaling of $z$, involving the 
Catalan generating function. Expanding \Dgammainf\ in $z$, we obtain 
\eqn\lime{e_\Gamma(x)\buildrel {\Gamma \to \infty}\over \to 
e(x)\equiv \left\{ \matrix{0 & \ \hbox{if $x$ odd} \cr {x!\over 2^{x+1} 
\left({x\over 2}\right)! 
\left({x\over 2}+1\right)!} & \ \hbox{if $x$ even} \cr }\right.}
If we admit that relative distances between degenerate minima are 
the appropriate statistical quantities to describe the distance between
two diffusing particles, this result without scaling of $x$ at large
times agrees with the idea of localization \Goldeux. Still, as explained
in \Lal , the distribution $e(x)$ behaves like $e(x) \sim x^{-3/2}$
at large $x$. 
Therefore, its moments $\langle x^\alpha \rangle$ diverge for $\alpha>1/2$.
This divergence comes precisely from the rare configurations (occurring
with probability $1/\Gamma$) with two minima at a distance of order $\Gamma^2$.
They contribute to $\langle x^\alpha \rangle$ by a term 
$(\Gamma^2)^\alpha/\Gamma = \Gamma^{2\alpha-1}$ which diverges at large 
$\Gamma$ for $\alpha>1/2$ (see \Lal ).

\subsec{Average position of the minimum }

{}From the formulas for the generating functions ${\cal P}_\Gamma^{(i)}(z)$, 
it is straightforward to get the average position 
$\overline{x}^{(i)}(\Gamma)$ and average squared position 
$\overline{x^2}^{(i)}(\Gamma)$ of the minimum for the three prescriptions:
\eqn\avpos{\eqalign{
\overline{x}^{(i)}(\Gamma)&=\sum_{x=0}^\infty x p_\Gamma^{(i)}(x)=
z\left.{d\ \over dz}\right\vert_{z=1} {\cal P}^{(i)}_\Gamma(z)\cr
\overline{x^2}^{(i)}(\Gamma)&=\sum_{x=0}^\infty x^2 p_\Gamma^{(i)}(x)=
\left.\left(z{d\ \over dz}\right)^2\right\vert_{z=1} 
{\cal P}^{(i)}_\Gamma(z)\cr
}}
Expanding \Ptwoform\ in powers of $z-1$ around $z=1$ (or in powers of $t$ 
around $t=0$), we get 
\eqn\xgamma{\eqalign{
\overline{x}^{(1)}(\Gamma)&={2\Gamma^2+5\Gamma+3\over 3}\cr
\overline{x}^{(2)}(\Gamma)&={2\Gamma^2+7\Gamma+3\over 3}\cr
\overline{x}^{(3)}(\Gamma)&={2\Gamma^2+6\Gamma+3\over 3}\cr}}
and
\eqn\xdeuxgamma{\eqalign{
\overline{x^2}^{(1)}(\Gamma)&= {16 \Gamma^4+88 \Gamma^3 +174 \Gamma^2
+147 \Gamma +45\over 15}\cr
\overline{x^2}^{(2)}(\Gamma)&= {16 \Gamma^4+104 \Gamma^3 +226 \Gamma^2
+179 \Gamma +45\over 15}\cr
\overline{x^2}^{(3)}(\Gamma)&= {16 \Gamma^4+96 \Gamma^3 +{1780\over 9} 
\Gamma^2 +163 \Gamma +45\over 15}\cr
}}
All of them lead to the same large $\Gamma$ asymptotic formulas
\eqn\asymres{\eqalign{
\overline{x}^{(i)}(\Gamma)& \buildrel {\Gamma\to \infty}  \over 
 \sim {2\over 3}\Gamma^2 \cr
\overline{x^2}^{(i)}(\Gamma)& \buildrel {\Gamma\to \infty}  \over
\sim {16\over 15}\Gamma^4 \cr}}
This asymptotic behavior involves only the scaling variable $x/\Gamma^2$
and the exact prefactors can be directly obtained from the limiting 
distribution \limdist\ by a suitable expansion in $s$ around $s=0$. 
We have in particular the asymptotic relation
\eqn\asymlaw{\overline{x^2}\sim {12\over 5} \left(\overline{x}\right)^2}
Here again, the difference for the subleading terms in \xgamma\ and
\xdeuxgamma\ for the three prescriptions comes from landscapes with 
two degeneracies separated by a distance of order $\Gamma^2$.
Such configurations, occuring with probability $\sim 1/\Gamma$, 
yield a correction of order $\Gamma^2/\Gamma=\Gamma$ 
to $\overline{x}(\Gamma)$ and a correction of order 
$\Gamma^4/\Gamma=\Gamma^3$ to $\overline{x^2}(\Gamma)$. For
these situations with exactly {\it two} effective degenerate minima, 
it is equivalent to take the average over the  minima (prescription (3)) 
and to average over the closest and the furthest minimum 
(prescriptions (1) and (2)). This explains why the subleading 
coefficient in the prescription (3) is exactly the average of the 
subleading coefficients of prescriptions (1) and (2). Situations with 
three degenerate minima distant from $\sim \Gamma^2$ occur with 
probability $\sim 1/\Gamma^2$ and influence the sub-subleading coefficient. 
Note also that the constant term in \xgamma\ and \xdeuxgamma\ must be the 
same for the three prescriptions since, for $\Gamma=0$, there cannot be any 
degeneracy of the deepest minimum. The above arguments explain why 
$\overline{x}^{(3)}(\Gamma)$ is exactly the average of 
$\overline{x}^{(1)}(\Gamma)$ and $\overline{x}^{(2)}(\Gamma)$ and
why  $\overline{x^2}^{(3)}(\Gamma)$ differs form the average of
$\overline{x^2}^{(1)}(\Gamma)$ and  $\overline{x^2}^{(2)}(\Gamma)$
by a term of order $\Gamma^2$ only.
\medskip 
In order to have a precise measure of the subleading terms, we
introduce the quantity
\eqn\ratiochi{\chi_\Gamma^{(i)}={\overline{x^2}^{(i)}(\Gamma)
\over \left({12\over 5}\right)\left(\overline{x}^{(i)}(\Gamma)\right)^2}
-1}
with a factor $12/5$ chosen to eliminate the leading large $\Gamma$
term, so that $\chi_\Gamma^{(i)}\sim 1/\Gamma$ tends to zero at large 
$\Gamma$. Remarkably, in the case (3), an extra cancelation occurs, 
leading to $\chi_\Gamma^{(3)}\sim 1/\Gamma^2$. 
In the following section, we compare this quantity, as computed
from \xgamma\ and \xdeuxgamma , with a similar quantity defined for 
the dynamical model.

\newsec{Dynamics}

\subsec{Simulation}

Beside the above theoretical statistical predictions, we have made
a numerical study of the dynamics \jum\ of a particle for a large 
sample of quenched random potentials drawn to satisfy the relations 
\height. 

For each drawn landscape, we calculate the probability
$P(x,t)$ that the particle sits at position $x$ at time $t$.
Due to the discrete nature of the landscape, it is possible
to make an {\it exact} enumeration of all the possible walks 
arriving at a given position $x$ at some time $t$, and to evaluate 
their probability deduced from \jum\ for the particular chosen
landscape. 

A more efficient way to implement this enumeration is to calculate 
$P(x,t)$ exactly step by step in time from the master equation
\eqn\master{P(x,t+1)=p_{x-1}P(x-1,t)+q_{x+1}P(x+1,t)}
with $p_x$ and $q_x=1-p_x$ the probabilities of jumping from site
$x$ respectively to the right and to the left, as explicited in
\jum\ , with in particular $p_0=1-q_0=1$ and with the convention
$p_{-1}=0$.  The master equation is supplemented by the initial 
condition $P(x,0)=\delta_{x,0}$. The computed probability $P(x,t)$ 
is then averaged over all the energy landscapes of our sample
(the sample size is $10^5$ energy landscapes for most simulations).

At any finite $t$, it is clear that the particle cannot reach a position 
$x>t$, hence $P(x,t)=0$ exactly for all $x>t$. The normalization of 
the probability requires that $\sum_x P(x,t)=1$ at any time $t$.  
In practice, the probability to occupy sites far from the origin is 
extremely low. To reduce the computation time, we drop the normalization
condition and replace it by
\eqn\sumrule{\sum_x P(x,t) > 1-\varepsilon}
where $\varepsilon\ll 1$. Therefore, instead of describing the complete 
accessible landscape of size $x=t$ at time $t$, we fix the maximal size 
at a much smaller value of $x$ such that the condition \sumrule\ is 
fulfilled for all our energy landscapes. Of course, the choice of this 
size depends crucially on the temperature $T$ and on the number of iterations.
In practice, we take $\varepsilon=0.01$ for most simulations and we checked 
that this simplification does not lead to significant errors. 
The computation time is significantly reduced for low temperatures, 
since the effectively visited landscape is of much smaller size. With this
simplification, we were able to study the dynamics up to $10^5$ iterations 
for low temperature regimes ($10^4$ for $T=2$).

Note that for our particular choice of dynamics, the particle cannot 
stay at the same site for two consecutive times $t$ and $t+1$. Thus
$P(2x+1,2t)=P(2x,2t+1)=0$. As we already mentioned, the net effect of this 
parity condition is to create residual fluctuations, which persist 
even at $T=0$, where barriers of height $\Gamma=1$ can always be passed. 
This will explain in particular why, at $T=0$, some equilibration can
take place between all the deepest minima accessible by passing 
$\Gamma=1$-barriers.
In the following subsections, we always present results for times of a well 
defined (even) parity. 

\fig{Evolution with time (in logarithmic scale) of the distribution $P(x,t)$ 
in a given energy landscape (drawn below). The evolution runs over $10^7$ 
iterations. The intensity in the greyscale is proportional to 
$-\ln P(x,t)$, i.e. darker regions
correspond to higher values of $P(x,t)$.}{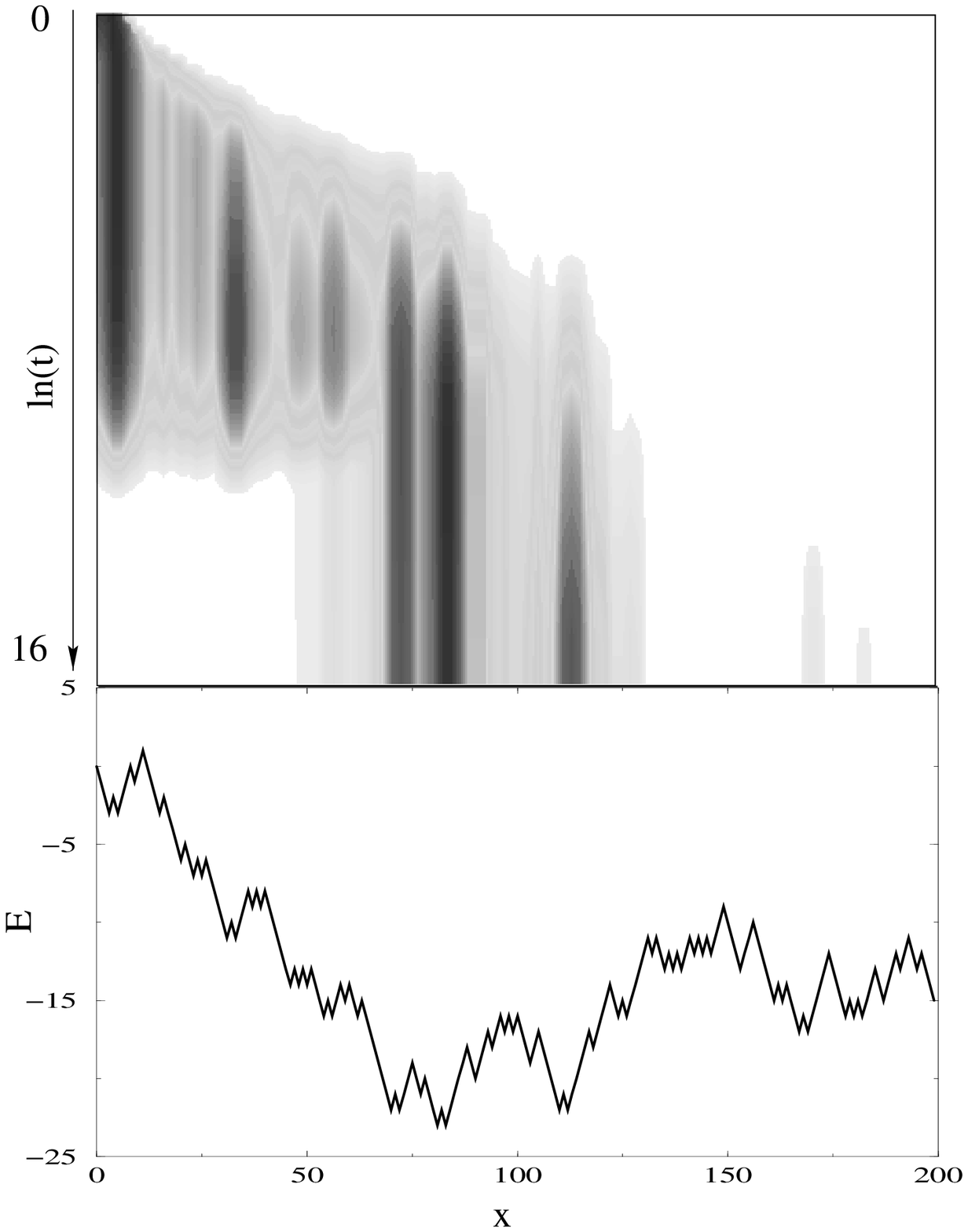}{11.truecm}
\figlabel\paysage

It is instructive to visualize the typical evolution of $P(x,t)$ with time for a 
fixed landscape before making the quenched average over our sample of landscapes.
Figure \paysage\ shows such an evolution with $10^7$ iterations. One clearly sees
that the regions with a large probability of occupation are concentrated around
local minima of the potential. As $t$ increases, these high density regions 
migrate 
to deeper minima. The duration of occupation of a local minimum in logarithmic
scale is roughly proportional to the height of the energy barrier on its right.

\subsec{Distribution P(x,t)}
 
Let us first present our numerical results for the distribution $P(x,t)$.
We will use overbars to denote the average over our sample of landscapes,
while brackets $\langle\cdot\rangle$ will denote the thermal
average estimated from the probability $P(x,t)$ computed for a fixed 
landscape. We are interested here in the average distribution
$\overline{P(x,t)}$ at a {\it fixed} large enough time.
{}From the asymptotic large $\Gamma$ results \probdens\ and \asymres , 
we expect that, at large $t$,
\eqn\asymdist{\overline{\langle x \rangle}(t)
\overline{P(x,t)}\sim {2\over 3}\ p_\infty\left({2\over 3}{x\over 
\overline{\langle x \rangle}(t)}\right)}
with $p_\infty$ given by \limdens. 
\fig{The rescaled average distribution $\overline{\langle x\rangle}(t)
\overline{P(x,t)}$ as a function of the rescaled variable 
$x/\overline{\langle x\rangle}(t)$ at $T=1/2$ and for various values of 
the time $t=1\,000$, $5\,000$, $10\,000$, $50\,000$ and $100\,000$. The solid
line indicates the exact asymptotic formula as given by 
\asymdist\ and \limdens .}
{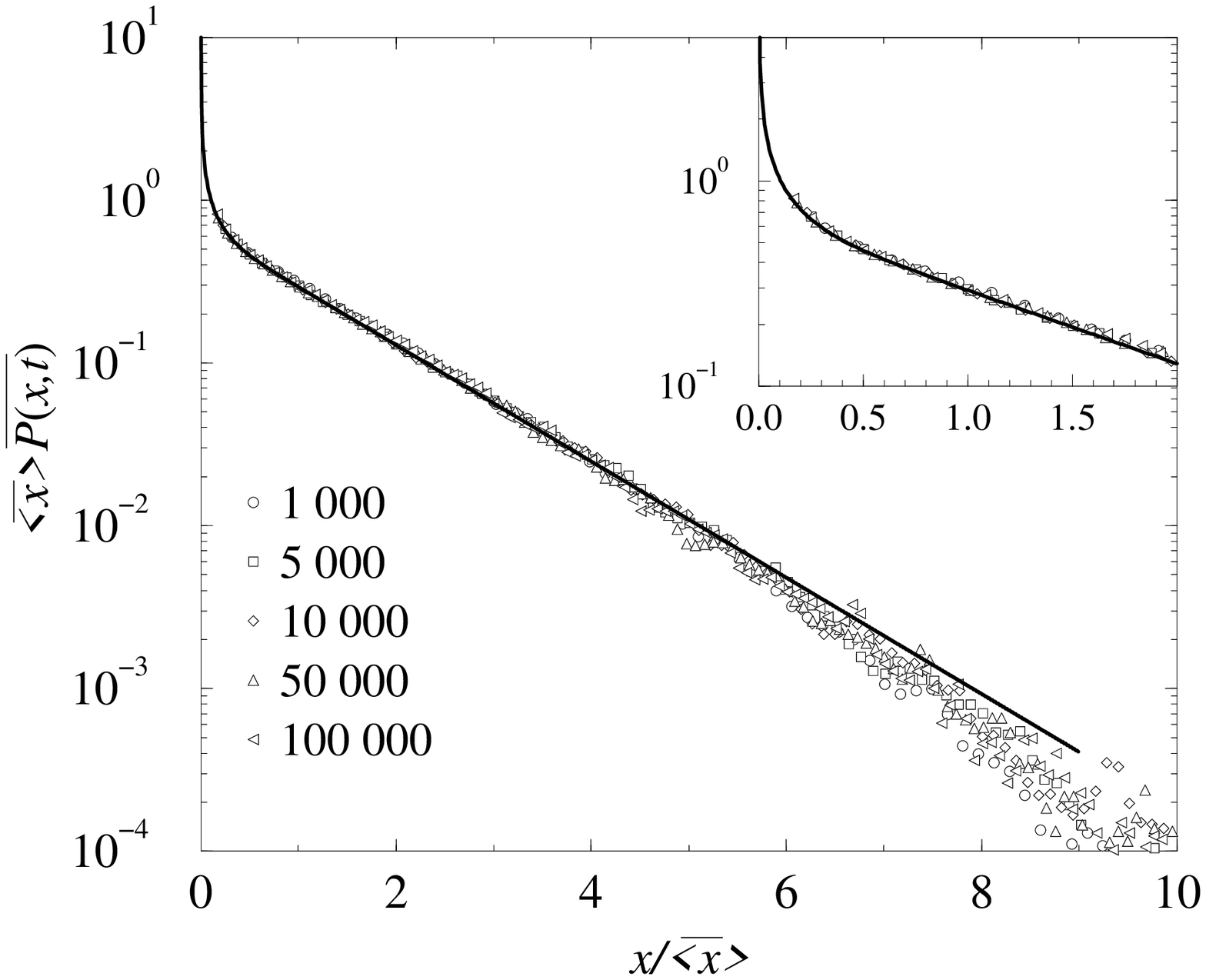}{10.cm}
\figlabel\distribution
Figure \distribution\ shows our results for $T=1/2$ and different 
values of $t$. The agreement with the asymptotic exact formula
is apparently very good. To have a better quantitative evaluation
of how close we are to the asymptotic result, we will study in the
next section the first and second moments of the distribution. 
As we shall see, significant deviations do actually exist, 
some of which can be well explained by our finite $\Gamma$ corrections
to the asymptotic statistics. We will also discover some interesting 
underlying oscillatory behaviors. 

\subsec{Results for the first and second moments and comparison with 
the statistics of minima}

We present here our numerical results for the average 
position $\overline{\langle x\rangle}(t)$ and average squared 
position $\overline{\langle x^2\rangle}(t)$ for varying time $t$
and at various temperatures.
We first check if $\overline{\langle x^2\rangle}(t)$ and
$(\overline{\langle x\rangle}(t))^2$ obey the expected asymptotic relation
\asymlaw\ with a proportionality factor $12/5$. Note that for a purely 
diffusive system in a homogeneous medium (i.e. in a flat energy landscape) 
where $\langle x\rangle(t) \sim \sqrt{t}$, a similar relation holds,
but with a smaller proportionality factor $\pi/2$. 
\fig{A numerical check of the relation \asymlaw . The solid line has the
slope $12/5$ expected from \asymlaw\ while the dashed line has the slope
$\pi/2$ expected for the normal diffusion is a flat landscape. For $T=2$, a
broader window of $\overline{\langle x\rangle}^2$ (see inset) is needed to 
recover the asymptotic law.}{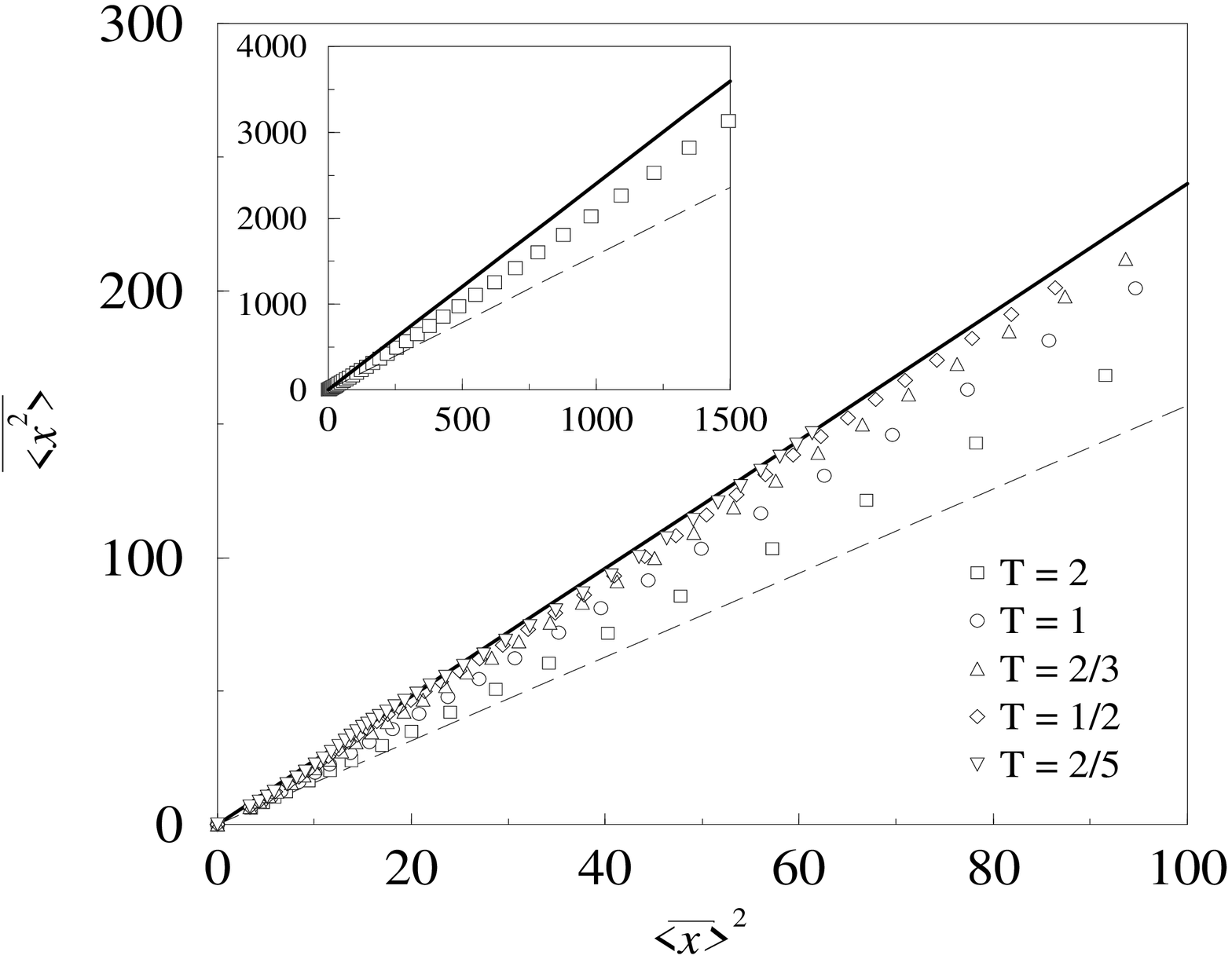}{10.cm}
\figlabel\xxdeux
Figure \xxdeux\ presents the corresponding data for several temperatures 
from $T=2/5$ to $T=2$. For the range of $\overline{\langle x\rangle}^2$ 
presented here, the asymptotic 
formula \asymlaw\ and the more complete statistical relations obtained 
from our finite $\Gamma$ predictions by eliminating $\Gamma$ between 
\xgamma\ and \xdeuxgamma\ do not differ significantly. We therefore 
expect, if the dynamics follows the statistics of minima, that the asymptotic 
linear relation is verified in the whole range of 
$\overline{\langle x\rangle}^2$. 
This is precisely what we observe at low temperatures (below $T\sim 1$). 
In this regime, a more refined analysis will reveal very interesting underlying
oscillatory behaviors, as emphasized below.
At high temperatures (above $T\sim 1$), we see a significant deviation from the
expected law, with a behavior closer to a purely diffusive regime at small 
$\overline{\langle x\rangle}^2$. Still, even at $T=2$ (see inset in Figure 
\xxdeux ), the correct slope is eventually recovered at large 
$\overline{\langle x\rangle}^2$. 
In this range of temperatures, the observed deviation is not explained by our finite 
$\Gamma$ corrections to the asymptotic law $p_\infty$, but is more simply the 
effect of a short time diffusion regime in which the particle does not yet feel 
the random potential. Such a regime will last until the particle reaches a
distance $x$ such that $E(x)\sim \sqrt{x}\sim T$, i.e up to a transition time 
$t\sim x^2\sim T^4$.
\medskip
For a comparison of our data with the statistics of minima of the previous 
section, we need to consider instead of the first and second moments more 
refined quantities which in practice contain exactly the same information 
but are more adequate for our purposes since they clearly emphasize the finite 
$\Gamma$ corrections. Anticipating our conclusions, we focus on the statistics 
$(3)$ which corresponds to an averaging over degenerate minima.
Inverting the formula \xgamma\ for the prescription $(3)$, we 
consider instead of $\overline{\langle x\rangle}(t)$ the equivalent
quantity:
\eqn\gammadet{\Gamma(t)\equiv {-3+\sqrt{3+6\overline{\langle x\rangle}(t)}
\over 2}}
such that $\overline{x}^{(3)}(\Gamma(t))=\overline{\langle x\rangle}(t)$.
The quantity $\Gamma(t)$ is thus an estimate of the effective height
of the barriers which can be passed at the time scale $t$, obtained by
matching the first moment measured in the dynamical process with 
the average position of the minima resulting from the formula \xgamma\ 
in case $(3)$. 

\fig{The quantity $\Gamma(t)$ as defined by Eq. \gammadet\ for $T=1/20$ to $T=2$,
as a function of $\ln(t)$. The solid lines show the corresponding fits of Eq. 
(4.4) for the average linear growth of the curves remaining after discarding
the superimposed oscillations. The use of the complete formula \gammadet\ is
necessary to obtain the correct position of the fit at low temperatures. At high 
temperature ($T=2$ in the inset), the fit (4.4) is better if we define 
$\Gamma(t)$ by the asymptotic formula $\Gamma(t)\sim \sqrt{(3/2)
\overline{\langle x\rangle}(t)}$
}{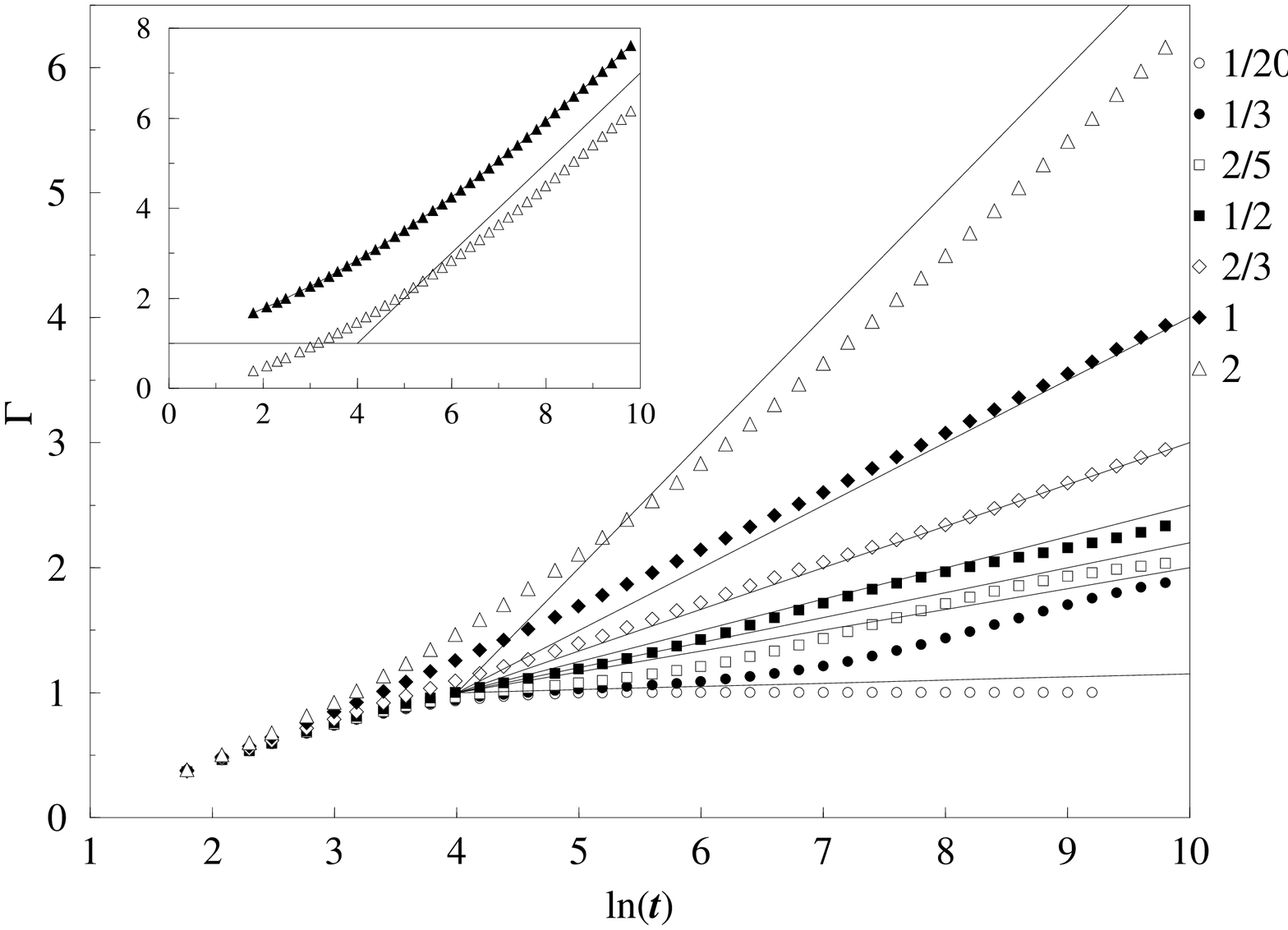}{10.cm}
\figlabel\gammavslnt

Figure \gammavslnt\ shows $\Gamma(t)$ as a function of ${\ln}(t)$ for different
values of the temperature. At low temperature, $\Gamma(t)$ oscillates around 
an average 
straight line and develops plateaus at {\it integer} values of $\Gamma$.
These plateaus are of course a signature of the underlying discrete nature
of the landscape, and are an indication of the actual relation between
the dynamics and a process of passing increasing discrete barriers.
This effect disappears at higher temperatures. Discarding these oscillations,
the curves have an average linear growth with a slope directly proportional to 
the temperature. More precisely, we can reasonably fit this average linear growth
by the formula (see Figure \gammavslnt : for the oscillating curves, the fits 
reasonably match the maxima of the oscillations)
\eqn\fitslope{\Gamma(t)={T\over 2}{\ln}\left({t\over t_0}\right)+1}
with ${\ln}(t_0)\sim 4$.
The proportionality factor $1/2=T/\sigma$ between $\Gamma(t)$ and 
$T{\ln}(t)$ is that expected from the correspondence between 
statistics and dynamics in the asymptotic limit, as already discussed. 
At low temperature and shorter times, this correspondence 
is still reasonably good, apart from the superimposed oscillations. 
The additive constant $1$ in \fitslope\ can be understood as the effect of 
the ``residual fluctuations" which remain at $T=0$ from the parity condition 
and make the barriers of height $\Gamma=1$ always passable. To obtain the 
correct position of the linear fit, we definitely had to use the complete 
formula \gammadet , 
which presents a shift of $-3/2$ with respect to the asymptotic relation 
$\Gamma(t)\sim \sqrt{(3/2)\overline{\langle x\rangle}(t)}$ obtained from 
\asymres . This reflects the importance of the finite $\Gamma$ corrections
at low temperatures. For higher temperature however ($T=2$), we obtain a better 
scaling with the asymptotic law without shift (see the inset in Figure 
\gammavslnt , black triangles) than with the shift (white triangles). 
Again at high temperatures,
the deviation from the asymptotic limit is not explained by finite $\Gamma$ 
corrections alone.
 
We now analyze our data for the second moment $\overline{\langle x^2\rangle}(t)$.
Here again, we prefer to consider the more adequate quantity
\eqn\chidet{\chi(t)={\overline{\langle x^2\rangle}(t)\over 
\left({12\over 5}\right)\left(\overline{\langle x\rangle}(t)\right)^2}-1}
copied from \ratiochi\ to measure the deviation from the asymptotic
regime \asymlaw . 

\fig{The quantity $\chi(t)$ defined by Eq. \chidet\ as a function of
$\Gamma(t)$ for temperatures ranging from $T=1/5$ to $T=2$. The solid
line shows the statistical estimate $\chi^{(3)}_\Gamma(\Gamma)$ as defined 
by Eq. \ratiochi .  The cross in the inset indicates the small corrections at 
$\Gamma=1$ to this statistical value due to parity effects, as computed
in Appendix A. The dashed line in the lower inset indicates the value of 
$\chi$ for pure diffusion in flat landscape.}{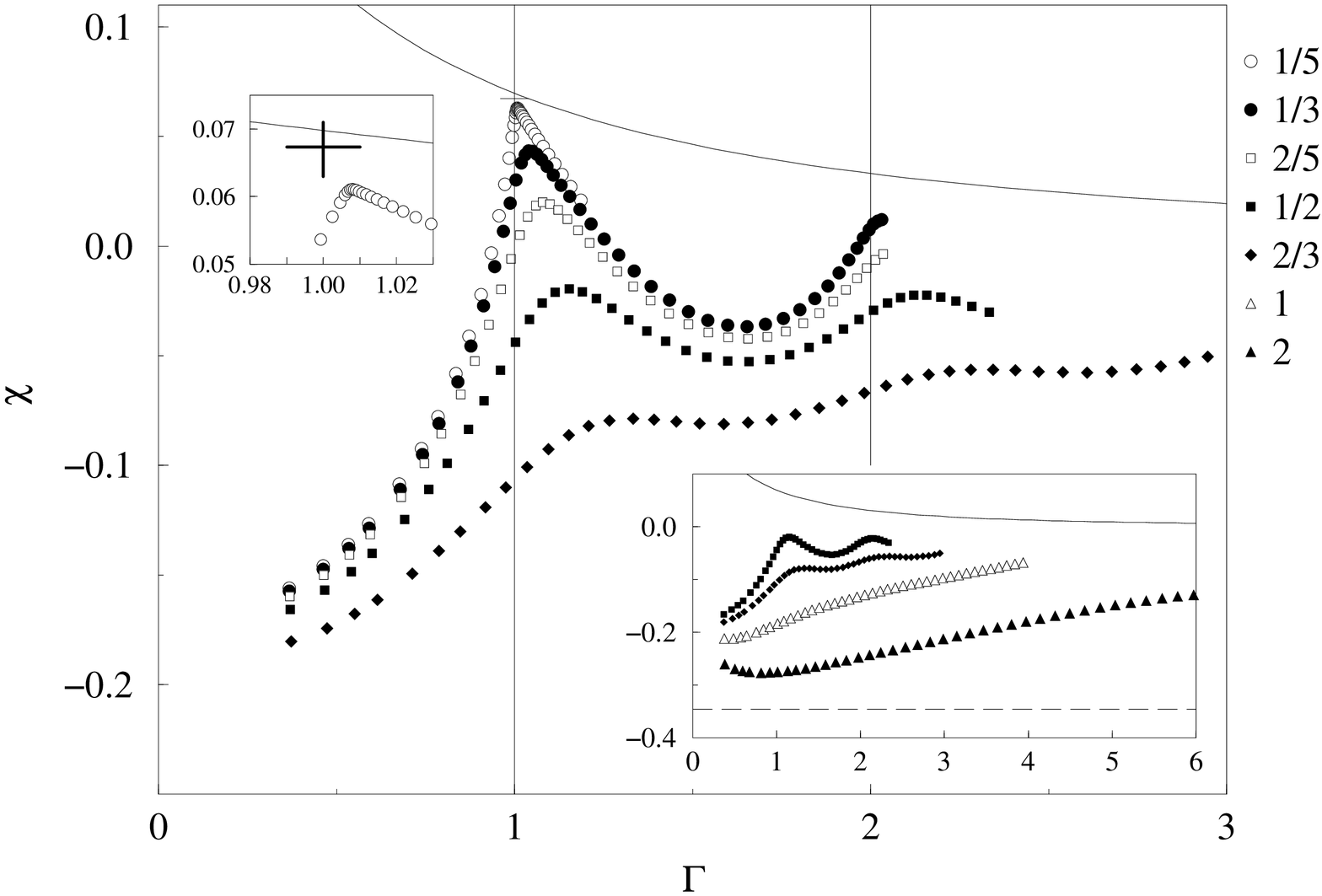}{10.cm}
\figlabel\chivsgamma 

Figure \chivsgamma\ presents our results for $\chi(t)$ versus $\Gamma(t)$,
and a comparison with the corresponding statistical relation calculated
from the previous section between $\chi_\Gamma^{(3)}$ and $\Gamma$.
We recover again a low temperature regime with oscillations and a high
temperature regime without oscillations. In this high temperature regime,
the asymptotic limit is reached very slowly and the short time dependence
is different from the finite $\Gamma$ predictions for the statistics
of minima. The subleading short time corrections to the universal asymptotic 
behavior are thus different from that predicted by the statistics of minima.
At low temperature however, the oscillations of $\chi$ are peaked around
integer values of $\Gamma$, and sharpen as the temperature decreases
to zero. In this limit $T\to 0$, the height of the peaks tend 
precisely to the value of $\chi_\Gamma^{(3)}$ at the corresponding
integer. We thus recover our predictions for the statistics of
minima, which strictly speaking are valid only for integer values
of the barrier $\Gamma$. We also observe two additional dynamical effects: 
\item{-} Between two consecutive integer values $\Gamma-1$ and $\Gamma$, 
we find a transition regime with a strong depletion of $\chi$. We
interpret this effect as resulting from a period of equilibration of the 
particle passing from the statistics of minima at scale $\Gamma-1$ to 
that at scale $\Gamma$. 
\item{-} As the temperature increases, the peaks are rounded and their maxima
slowly move to lower values of $\chi$. The peaks eventually 
disappear at high temperature.

We interpret the above results as follows. {}From the formula
\fitslope , the barriers of height $\Gamma$ are passed at
times of order $t(\Gamma)\sim t_0\exp[2(\Gamma-1)/T]$. After passing these
barriers, we admit that the time needed for equilibration in the
(always present) degenerate minima is itself of the order of a multiple 
of $t(\Gamma)$. As the temperature is lowered, the corresponding
proportionality factor remains finite due to the residual fluctuations. 
During the equilibration process at a given $\Gamma$, the data get closer 
to the equilibrium distribution 
of the minima. The time needed to pass the next 
barriers, i.e. those of height $\Gamma+1$ is from \fitslope\ 
$t(\Gamma+1)=t(\Gamma)\exp(2/T)$, that is again a finite 
multiple of $t(\Gamma)$, depending only on the temperature. In the low 
temperature regime, the particle thus has enough time to equilibrate and
recover the statistics of minima for a fixed passable height $\Gamma$
{\it before} it starts passing the barriers of height $\Gamma+1$.
Conversely, at high temperature, the particle keeps finding better and
better minima by passing increasing barriers without equilibration 
for each passed $\Gamma$. In particular, the particle does not feel the
discrete nature of the potential. This results in the suppression of the 
oscillations and a behavior closer to normal diffusion.
\medskip
In the above analysis, the use of statistics $(3)$ is crucial on the one hand
to get peaks precisely at integer values of $\Gamma(t)$ and on the other hand 
to recover the theoretical value of $\chi(\Gamma)$ at the peak for low 
temperatures.
These conditions eliminate the two other statistics. At this level of 
precision, the purely asymptotic result $\chi=0$ is also ruled out.
As discussed in the Appendix A, the statistics $(3)$ itself must be modified by 
very small corrections due to parity effects, i.e. the fact that
the particle at time $t$ cannot sit right at the correct minima if those 
happen to have a parity different from $t$. This effect is sensible only 
for the peak at $\Gamma=1$. The corresponding correction is calculated in 
the Appendix A and leads to a very small reduction ($\sim 3.5\%$) of 
the peak (see Figure \chivsgamma ), consistent with the data of the dynamical 
process. The parity correction is more important for quantities which
measure the localization of the particle, as we discuss in the next section.
The use of a different dynamics allowing the particle to remain at the same site 
should suppress this correction. 

\subsec{Localization}

In order to have an idea of how localized is the particle, we have 
measured the probability second moment:
\eqn\ydeux{Y_2(t)=\overline{\sum_x \left(P(x,t)\right)^2}}
which estimates the probability that two independent particles
evolving in the same quenched potential arrive {\it exactly}
at the same site at time $t$. This quantity is similar to the participation
ratio for localized quantum particles \Compte. A non zero value of $Y_2$
is the signal of a localization. We must however distinguish between 
two different effects which run counter to this localization and 
therefore lower
$Y_2$: the effect of temperature which broadens the distribution of a
particle around its average position in a minimum, and the existence of 
several degenerate minima in which the particles can fall. It is this
last effect that we measure at very low temperatures. If we assume
that, when $T\to 0$, the particle is localized exactly in the deepest
minima at scale $\Gamma(t)$, we estimate $Y_2(t)$ by
\eqn\estimyd{\eqalign{Y_2(\Gamma)&=\sum_{k=0}^\infty{1\over k+1} 
{\Gamma+2\over 2(\Gamma+1)}\left({\Gamma\over 2(\Gamma+1)}
\right)^k\cr &={\Gamma+2\over \Gamma}\ln\left({2(\Gamma+1)\over
\Gamma+2}\right)\cr}}
with $\Gamma=\Gamma(t)$. In the formula above, we used the probability
\probdeg\ to have exactly $k+1$ degenerate minima, weighed
by the probability $1/(k+1)$ to have the two particles in the same
minimum. In practice, as we already noticed, 
the formula \estimyd\ must be corrected to account for the parity 
effects of the dynamics. At large $\Gamma$, this simply amounts to
reduce $Y_2$ by a factor $(\Gamma+2)/(2\Gamma+3)+(1/2)\times(\Gamma+1)
/(2\Gamma+3)$ since $Y_2$ must typically be divided by $2$ if the minima 
have the wrong parity (which occurs with probability $(\Gamma+1)/(2\Gamma+3)$ 
for even times, see Appendix B) and the particle has to sit on the two 
neighbors of the minimum. 
With this estimate, $Y_2(t)$ thus tends at large times 
($\Gamma(t)\to \infty$) to a finite value $(1/2)\ln 2$. 
The same formula \estimyd\ with the above reduction factor gives also a good 
estimate at small $\Gamma$, with for instance a value $Y_2\sim (12/5)\ln(4/3)
\sim 0.69$ at $\Gamma=1$.
In Appendix B, we present a detailed analysis of the exact corrections 
for the statistical model at $\Gamma=1$, leading to a value $Y_2\sim 0.706415$. 

\fig{The quantity $Y_2(t)$ as defined in Eq. \ydeux\ as a function of 
$\Gamma(t)$ for temperatures
ranging from $T=1/5$ to $T=2$. The horizontal bar is the value of the peak
estimated in Appendix B from the statistics with parity 
corrections.}{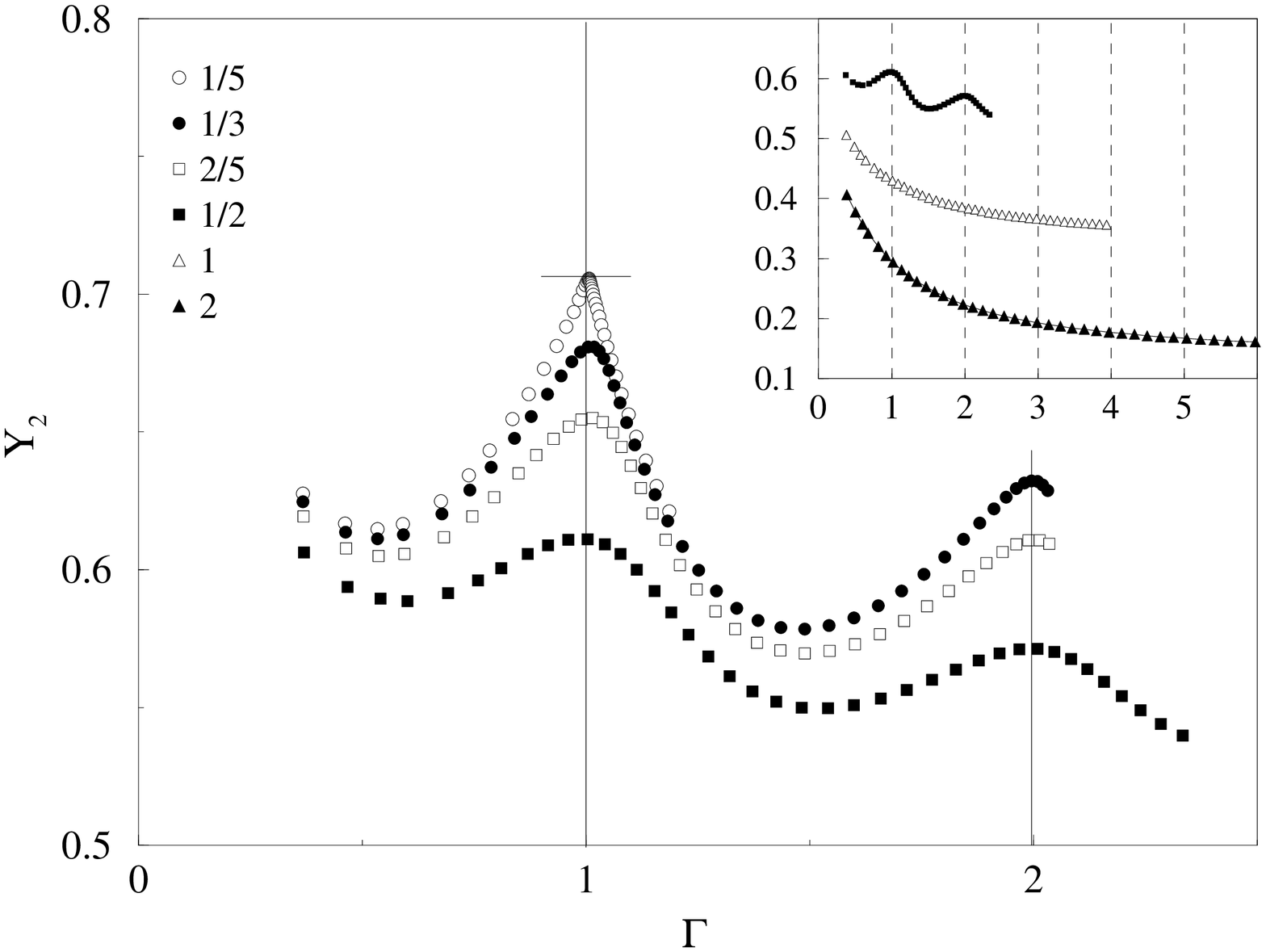}{10.cm}
\figlabel\ydeuxvsgamma

Figure \ydeuxvsgamma\ shows our numerical results for $Y_2(t)$ as a function
of $\Gamma(t)$. At low temperatures, we recover peaks at integer values of
$\Gamma$. At low temperatures, the value of the peak at $\Gamma=1$ is in 
perfect agreement with the estimate of Appendix B and our results are
consistent with a localization of the particle in all the degenerate
minima. At high temperatures, $Y_2(t)$ decreases with time but still
tends to a finite value at large $\Gamma$, apparently proportional
to $1/T$. We have also measured the more usual R\'enyi entropy
$H_2(t)=-\overline{\ln (\sum_x P^2(x,t))}$ \Gras . In contrast with $Y_2$, 
which is an average value over the disorder, $\exp[-H_2]$ gives the 
{\it typical} value of the probability for two particles to be at the same site
at time $t$. We find that $\exp[-H_2]$ and $Y_2$ display the same
behavior and differ by a roughly constant multiplicative factor.

Another quantity of interest for the measure of the localization of a
particle is the dispersion, defined by
\eqn\dispersion{\Delta x^2(t)=\overline{\langle x^2\rangle (t)
- \left(\langle x \rangle (t)\right)^2}}
If the particle were localized in a single minimum, the
dispersion $\Delta x^2(t)$ would not grow indefinitely with time
(or with $\Gamma$) but rather would reach a finite limit. 
The situation is quite different if, as we expect, the particle is 
localized in several degenerate minima since, as we already mentioned, 
minima can be separated by a distance of order $\Gamma^2$ with a 
probability $1/\Gamma$. In this case, $\Delta x^2$ grows like 
$(\Gamma^2)^2/\Gamma=\Gamma^3$ and is thus infinite asymptotically.
\fig{The quantity $\Delta x^2(t)$ (in logarithmic scale) as defined in Eq. 
\dispersion\ as a function of $\Gamma(t)$ (in logarithmic scale) 
for temperatures 
ranging from $T=1/10$ to $T=2$. The horizontal bar indicates the value of the 
peak estimated in Appendix B from the statistics 
with parity corrections.}{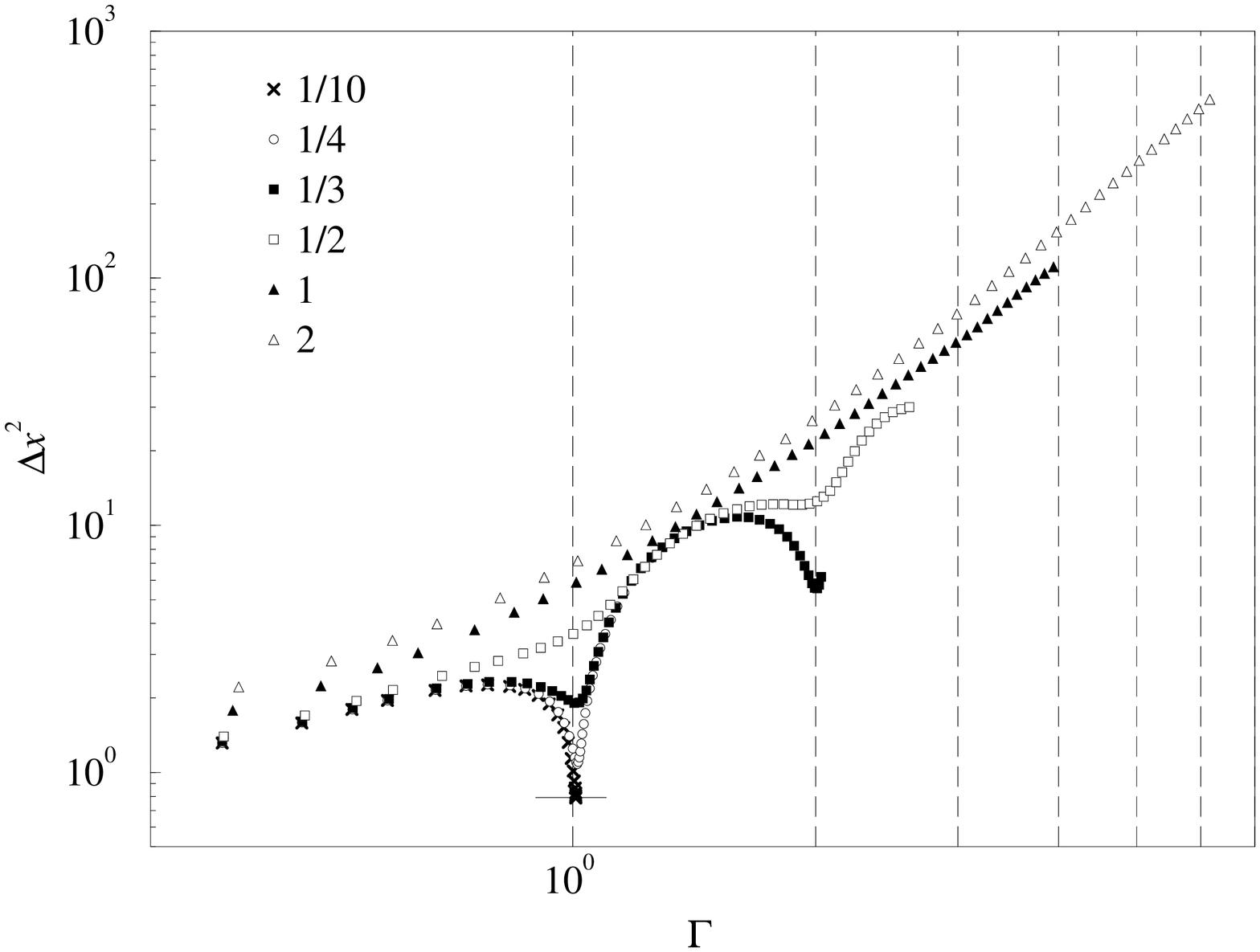}{10.cm}
\figlabel\dispvsgamma
Figure \dispvsgamma\ shows our numerical results for $\Delta x^2(t)$ as a 
function
of $\Gamma(t)$ in logarithmic scales. At low temperatures, the peaks
of $\Delta x^2(t)$ follow the statistics $(3)$ of the degenerate minima,
with the additional parity corrections, as computed for $\Gamma=1$ in
the appendix B. We thus recover a regime of localization in all the
degenerate minima. At high temperatures, $\Delta x^2(t)$ increases rapidly
with time, with a scaling compatible with the expected $\Gamma^3$
dependence. 

\newsec{Conclusions and discussion}

We have studied the problem of the one dimensional diffusion
of a particle in a semi-infinite quenched random energy landscape with
{\it discrete} integer heights and for different temperatures $T$. 
Our data converge to the expected large time universal asymptotic limit, 
but very slowly (as $1/\ln t$) and the approach to this limit crucially 
depends on temperature. To quantify the finite time corrections, we have 
compared our numerical data for the diffusion process at time $t$ 
with exact results for the statistics of the local minima of the landscapes 
reached by passing energy barriers of increasing (integer) size $\Gamma$. 
Below $T\sim 1$, we find a low temperature regime in which the dynamical
process precisely follows the statistics of minima which corresponds
to average over all the degenerate minima at a given scale $\Gamma$ with
the expected correspondence $\Gamma\leftrightarrow T\ln t$. Our data
are consistent with a localization of the particle equally distributed in all 
these degenerate minima. Interesting transition regimes interpolate between 
integer values of the scale $\Gamma$. 
At higher temperatures above $T\sim 1$, the approach to the
asymptotic laws do not follow the finite $\Gamma$ corrections
to the statistical laws, at least in the regime of times that we consider 
in our simulation. 

We interpret these two different regimes as follows:
at small $T$, the particle has enough time to equilibrate and recovers the
statistics at a given $\Gamma$ before it starts passing barriers of heights
$\Gamma+1$. At high temperatures however, the particle keeps passing barriers
of increasing $\Gamma$ without having enough time for equilibration at a fixed
$\Gamma$. 

A natural question is whether the existence of two regimes is only a finite 
time effect, or whether it persists for larger times. In the first case, 
the correspondence
between the dynamics and the statistics should be recovered at larger times. 
In the second case, it is tempting to expect a transition 
temperature between the two regimes.
The low temperature regime would be anyway a consequence of an 
underlying discrete cut-off for the steps of the energy landscape. 
The transition temperature would thus be of the order of this cut-off and 
would tend to zero in the continuum limit where only the high temperature 
regime would persist. 
\bigskip
\noindent{\bf Acknowledgments}

We thank C\'ecile Monthus for many useful discussions and for comments during
the redaction of the manuscript. We are also grateful to Andrea Baldassarri 
for interesting discussions. We finally thank Jean-Marc Luck for a critical 
reading of the manuscript.
 
\appendix{ A}{Corrections due to residual fluctuations at $T=0$}

We will concentrate here on our statistics $(3)$, which consists in averaging 
over all the degenerate minima with equal probability, that is, in the case of 
$k$ degenerate minima, in assigning a weight factor $1/k$ to each minimum. 
This statistics implicitly assumes that, after a transition period of 
equilibration, each degenerate minimum is visited with even probability. 

For our particular choice of dynamics however, the particle is not allowed 
to remain at the same site for two consecutive times $t$ and $t+1$. As we
already mentioned, this results in a parity condition, namely that the particle 
occupies even sites at even times and odd sites at odd times. 
Thus, even at $T=0$, the particle has to fluctuate from odd to even sites.
This is what we called ``residual fluctuations", which in particular 
make barriers of height $\Gamma=1$ always passable, even at $T=0$, 
eventually leading to some equilibration between all the deepest minima 
accessible by passing these $\Gamma=1$-barriers.

Another effect of these residual fluctuations is that, by preventing
the particle to remain seated right in the minima for all times,
they create small corrections to the estimates of statistics $(3)$, 
in particular for the limiting values of the peaks at $T\to 0$ of the 
quantities $\chi$, $Y_2$ and $\Delta x^2$. 

As we will now discuss, these corrections are of two types: a parity 
correction and a wall correction.

\noindent{\it Parity correction:} Let us consider a time $t$ with a given parity,
say even, so that the particle is forced to occupy sites with the same 
parity. For a fixed landscape and a fixed $\Gamma$, it is clear that all the 
degenerate deepest minima have a well defined, common parity 
since returning to the 
same height requires an even number of steps. This parity however may or may 
not be that of $t$. If the two parities coincide, then the particle can fall 
at $T\to 0$ precisely in the minima and our calculations using statistics $(3)$
are valid. If the two parities do not match, the particle cannot fall right 
in the deepest minima, but will rather occupy the two neighboring
positions on both sides 
of each minimum.  In this case, rather than averaging the positions 
$x_{\rm min}^{(m)}$, $m=1,\ldots,k$ of the $k$ minima with weights $1/k$, we 
should average the positions $x_{\rm min}^{(m)}\pm 1$ with weights $1/(2k)$ 
(of course, if two minima are distant by two elementary steps only, the point 
in between should receive a weight $1/(2k)+1/(2k)=1/k$). 
What is the probability for the minima to have the wrong parity? 
Looking at, say, the closest minimum, we can select the even minima by 
considering the combination 
$({\cal P}_\Gamma^{(1)}(z)+{\cal P}_\Gamma^{(1)}(-z))/2$ and the odd minima 
by considering $({\cal P}_\Gamma^{(1)}(z)- {\cal P}_\Gamma^{(1)}(-z))/2$ 
instead. 
The probability for a minimum to be even or odd is thus
\eqn\probevenodd{\eqalign{\hbox{Proba(even minima)}&={1\over 2} \left(
{\cal P}_\Gamma^{(1)}(1)+{\cal P}_\Gamma^{(1)}(-1)\right)={\Gamma 
+2\over 2\Gamma +3}
\cr \hbox{Proba(odd minima)}&={1\over 2} \left({\cal P}_\Gamma^{(1)}(1)
-{\cal P}_\Gamma^{(1)}
(-1)\right)={\Gamma +1\over 2\Gamma +3}\cr}}
As far as $\overline{x}$ is concerned, it is clear the parity shift creates no 
correction since $x_{\rm min}^{(m)}$ is precisely the average of 
$x_{\rm min}^{(m)}-1$ and $x_{\rm min}^{(m)}+1$. For $\overline{x^2}$ however, 
it results in a shift by a factor:
\eqn\shift{\sum_{m=1}^k{1\over 2k} \left((x_{\rm min}^{(m)}-1)^2+
(x_{\rm min}^{(m)}+1)^2\right)-{1\over k} (x_{\rm min}^{(m)})^2=1}
irrespectively of the number of degenerate minima. For even times, we thus get a
parity correction to $\overline{x^2}$ equal to 
$1\times (\Gamma+1)/(2\Gamma +3)$. 
This correction is negligible at large $\Gamma$ but can be 
measured at $\Gamma=1$,
where it predicts a shift of $2/5$ to $\overline{x^2}^{(3)}(1)$. 
The same effect is more sensible for $Y_2$ and $\Delta x^2$, 
and will be discussed in the Appendix B.

\noindent{\it Wall correction:} Another correction comes from the presence of the
wall at $x=0$ in the particular case where $x_{\rm min}=0$ itself is a minimum. 
This situation occurs with probability: 
\eqn\minzero{{\cal P}_\Gamma^{(1)}(0)={1\over \Gamma+2}}
Let us thus assume that $x_{\rm min}^{(1)}=0$ is the first deepest minimum, 
together 
with $(k-1)$ other minima at even positions $x_{\rm min}^{(m)}$, $m=2,\ldots k$. 
Let us also assume that $t$ is even so that the particle can sit 
precisely in these minima. Still, after equilibration, the minimum 
at $x_{\min}^{(1)}=0$ is less probable
that the other minima. This effect is visible at $T=0$ and $\Gamma=1$ where the
equilibration is due to residual fluctuations only. 
\fig{The modified equilibrium weights for $k$ degenerate minima at $\Gamma=1$
in the case of a first minimum right at the wall. Each weight at 
even times is the average of the neighboring weights at odd times and 
conversely.}{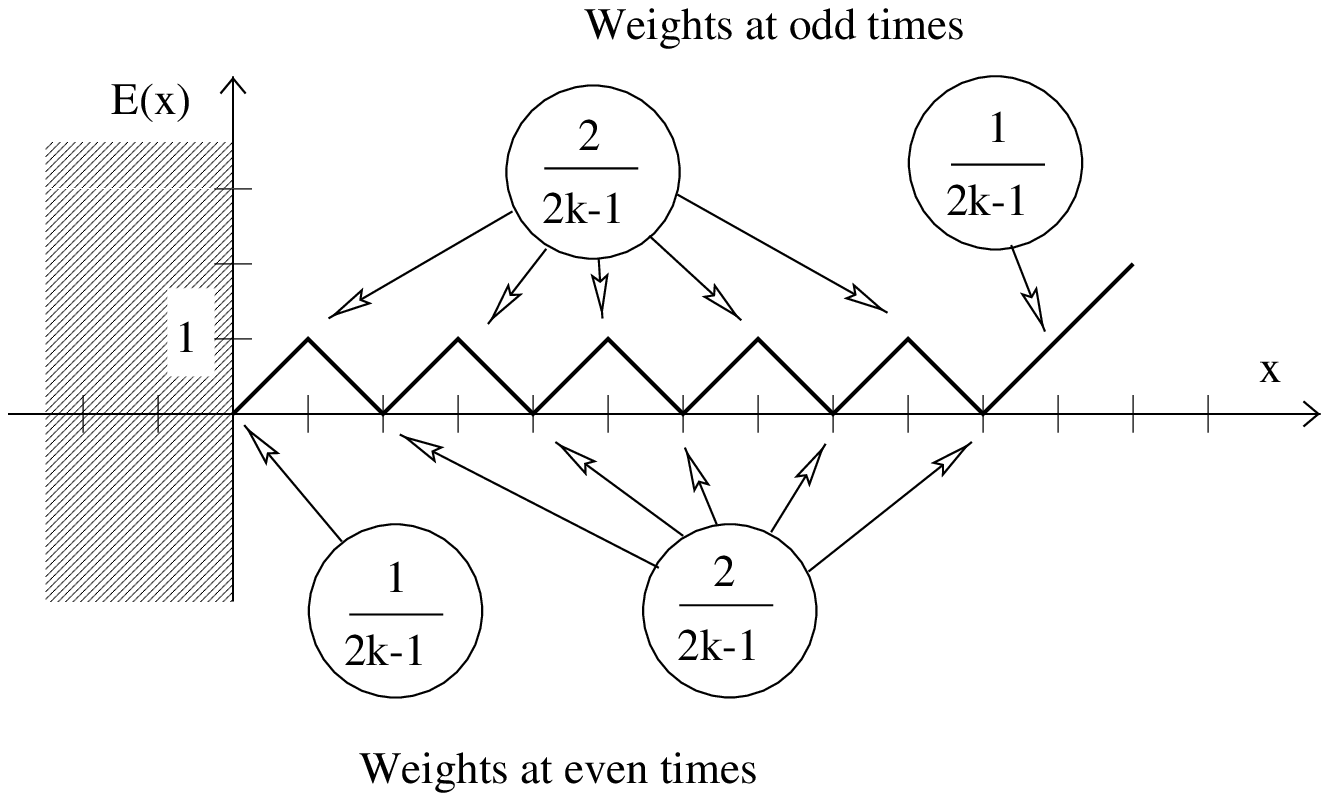}{8.cm}
\figlabel\wallcorrection
In this case (see Figure \wallcorrection), the $k$ minima are at positions 
$x_{\rm min}^{(m)}=2m-2$. Since the minimum at $x_{\rm min}^{(1)}$ has no
accessible neighbor on its left, it is easy to see that, at even times,
 this minimum is less probable that the others by a factor of $2$, leading to 
a probability $1/(2k-1)$ for this minimum and a probability $2/(2k-1)$ for 
the $(k-1)$ others, instead of an equal probability $1/k$ for each minimum.
The correction to $\overline{x}^{(3)}(1)$ is thus:
\eqn\corrx{\left({2\over 2k-1}-{1\over k}\right)\sum_{m=2}^k(2m-2)={k-1\over 
2k-1}}
and that to  $\overline{x^2}^{(3)}(1)$:
\eqn\corrxdeux{\left({2\over 2k-1}-{1\over k}\right)\sum_{m=2}^k(2m-2)^2=
{2\over 3} (k-1)}
According to \probdeg\ and \minzero , such a situation occurs with
probability $(1/3)\times (3/4)(1/4)^{k-1}$

Combining the parity correction and the wall correction, we get
\eqn\rescorrx{\overline{x}_{\rm corr.}=\overline{x}^{(3)}(1)+
{1\over 3}\times \sum_{k=1}^\infty {3\over 4}\left({1\over 4}\right)^{k-1}
{k-1\over 2k-1}= {92-3{\ln}(3)\over 24}}
instead of $\overline{x}^{(3)}(1)=11/3$, i.e. numerically $3.696$ instead
of $3.666$, and
\eqn\rescorrx{\overline{x^2}_{\rm corr.}=\overline{x^2}^{(3)}(1)+
{2\over 5}\times 1+{1\over 3}\times \sum_{k=1}^\infty{3\over 4}\left({1\over 4}
\right)^{k-1} {2\over 3}(k-1)= {4724\over 135}}
instead of $\overline{x^2}^{(3)}(1)=932/27$, i.e. numerically $34.99$ instead
of $34.52$. 
With these values, the estimate $\chi_1^{(3)}=0.06978$ is modified into
$\chi_{\rm corr.}=0.06733$, i.e. is lowered by $3.5\%$ only.

\appendix{ B}{Computation of $Y_2$ and $\Delta x^2$ in the limit $T\to 0$}

\fig{Respective weights of accessible minima at even times for
$\Gamma=1$ in the three situations: (i) the first minimum is right at 
the wall; (ii) the first minimum is at an odd position; (iii) the
first minimum is at an even position but not at the wall.}{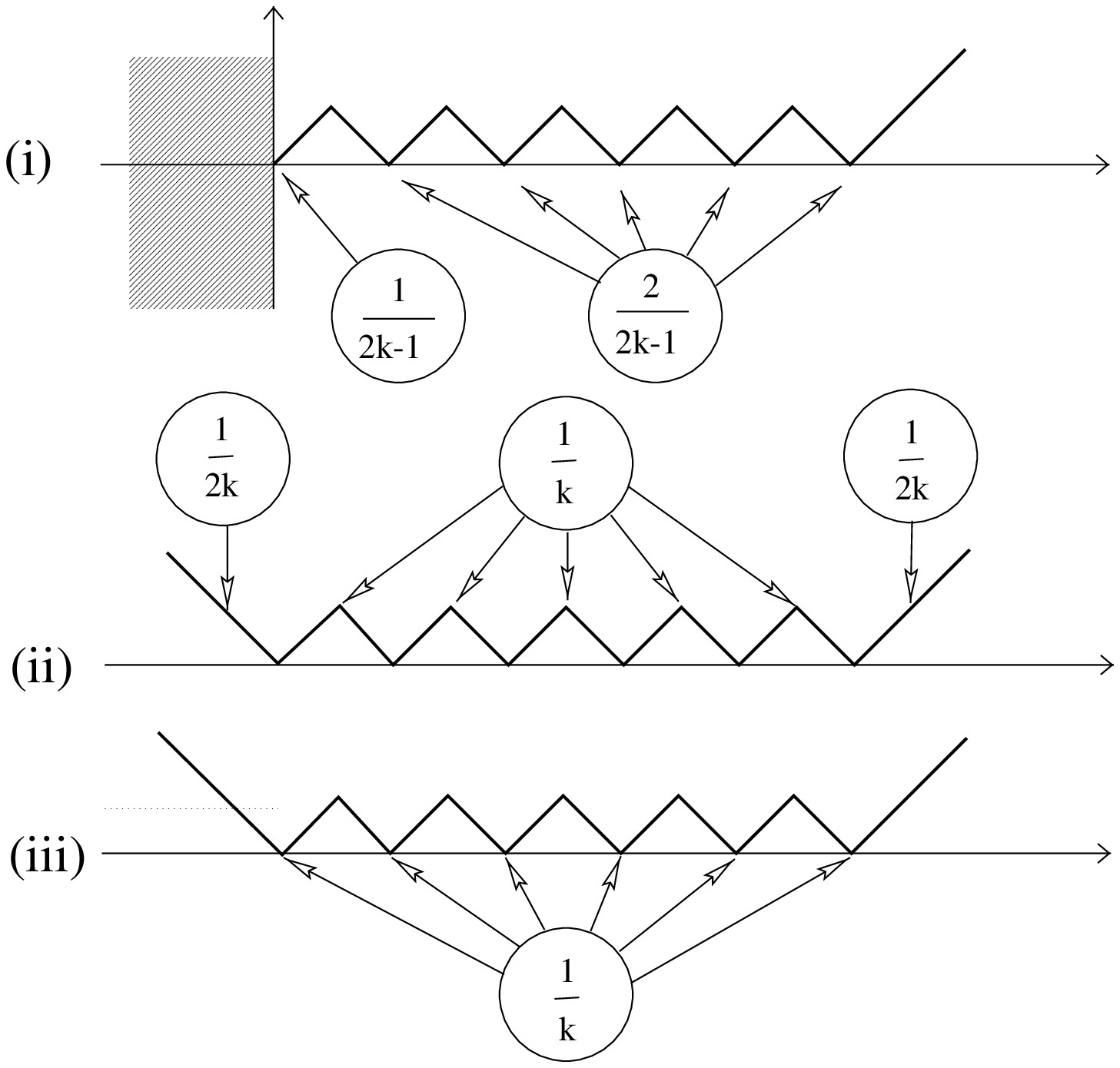}{9.cm}
\figlabel\threecases
In this Appendix, we will calculate the value of the peak at $T\to 0$
and $\Gamma=1$ of $Y_2$ and $\Delta x^2$. These value are very sensitive 
to the residual fluctuations. We will thus follow the same lines
as in the previous Appendix. 
We will consider even times and a situation with $k$ degenerate minima
at positions $x_{\rm min}^{(m)}=x_{\rm min}^{(1)}+2m-2$, $m=1,\ldots,k$.
According to \probdeg , this situation occurs 
with probability $(3/4)(1/4)^{k-1}$.
We distinguish between three possibilities for the first minimum:
\item{(i)} $x_{\rm min}^{(1)}=0$. According to 
\minzero , this occurs with probability $1/3$.
\item{(ii)} $x_{\rm min}^{(1)}$ is odd. According to \probevenodd ,this
occurs with probability $2/5$.
\item{(iii)} $x_{\rm min}^{(1)}$ is even and non zero. This
occurs with probability $1-1/3-2/5=4/15$.
\par
The weights associated to the minima (or their neighbors) in the three
situations are depicted in Figure \threecases .
The corresponding contributions to $Y_2$ are 
\eqn\contydeux{\eqalign{
 \left({1\over 2k-1}\right)^2+(k-1)\left({2\over 2k-1}\right)^2
&={4k-3\over (2k-1)^2}\quad \hbox{ in case (i)}\cr 
 \left({1\over 2k}\right)^2+(k-1)\left({1\over k}\right)^2
+ \left({1\over 2k}\right)^2 &={2k-1\over 2k^2}\quad \hbox{ in case (ii)}\cr 
 k\left({1\over k}\right)^2&={1\over k}\quad \hbox{ in case (iii)}\cr}}
Combining these contributions with their respective weights, we get
\eqn\resydeux{\eqalign{Y_2&=\sum_{k=1}^\infty{3\over 4}
\left({1\over 4}\right)^{k-1}
\left\{{1\over 3}\times {4k-3\over (2k-1)^2} +{2\over 5}\times 
{2k-1\over 2k^2}+{4\over 15}\times {1\over k}\right\}\cr
&=0.706415\cr}}

Similarly, the contributions to $\Delta x^2$ are 
\eqn\contdisp{
{2\over 2k-1}\sum_{m=2}^k(2m-2)^2-\left({2\over 2k-1}\sum_{m=2}^k(2m-2)\right)^2
={4k(k^3-2k^2+2k-1)\over 3(2k-1)^2}\quad \hbox{ in case (i)}}
\eqn\contdispdeux{\eqalign{ 
&{1\over 2k}(x_{\rm min}^{(1)}-1)^2+{1\over k}\sum_{m=2}^k(x_{\rm min}^{(1)}
+2m-3)^2+{1\over 2k}(x_{\rm min}^{(1)}+2k-1)^2
\cr &-\left({1\over 2k}(x_{\rm min}^{(1)}-1)
+{1\over k}\sum_{m=2}^k(x_{\rm min}^{(1)}
+2m-3)+{1\over 2k}(x_{\rm min}^{(1)}+2k-1)\right)^2
={k^2+2\over 3}\quad \hbox{ in case (ii)}\cr}}
\eqn\contdisptrois{ 
{1\over k}\sum_{m=1}^k(x_{\rm min}^{(1)}+2m-2)^2-\left({1\over k}\sum_{m=1}^k
(x_{\rm min}^{(1)}+2m-2)\right)^2
={k^2-1\over 3}\quad \hbox{ in case (iii)}}
Combining these contributions with their respective weights, we get
\eqn\resdisp{\eqalign{\Delta x^2&=\sum_{k=1}^\infty{3\over 4}\left({1\over 4}
\right)^{k-1} \left\{{1\over 3}\times {4k(k^3-2k^2+2k-1)\over 3(2k-1)^2} 
+{2\over 5}\times {k^2+2\over 3}+{4\over 15}\times {k^2-1\over 3}\right\}\cr
&=0.789288\cr}}

\listrefs
 
\end